\documentclass{article}

\usepackage{arxiv}

\usepackage[utf8]{inputenc} 
\usepackage[T1]{fontenc}    
\usepackage{hyperref}       
\usepackage{url}            
\usepackage{booktabs}       
\usepackage{amsfonts}       
\usepackage{nicefrac}       
\usepackage{microtype}      
\usepackage{lipsum}

\usepackage{physics} 
\usepackage{siunitx} 
\usepackage{enumerate} 
\usepackage{subfigure}
\usepackage{multirow}
\usepackage{float,lscape}
\usepackage{pgfplots}
\usepackage{pgfplotstable}
\usepackage{tikz,pgfplots}
\usepackage{amsmath}  
\usepackage{amssymb}
\usepackage[misc,geometry]{ifsym} 
\usepackage{footmisc}
\usepackage{doi}

\title{Do altmetrics work for assessing research quality?}

\author{
Andrea Giovanni Nuzzolese \thanks{Andrea Giovanni Nuzzolese is the main contributor of this paper and the principal investigator of the project that supported the research presented in this paper. Paolo Ciancarini, Aldo Gangemi, Silvio Peroni, Francesco Poggi, and Valentina Presutti contributed equally to this paper.}\\
Semantic Technologies Laboratory\\
Institute of Cognitive Sciences and Technologies\\
Italian National Research Council\\
Via San Martino della Battaglia 44, 00185 Rome, Italy\\
\texttt{andrea.nuzzolese@istc.cnr.it}
\And
Paolo Ciancarini\\
Digital and Semantic Publishing Laboratory\\
Department of Computer Science and Engineering\\
University of Bologna\\
Mura Anteo Zamboni 7, 00126 Bologna, Italy\\
\texttt{paolo.ciancarini@unibo.it}
\And
Aldo Gangemi\\
Digital Humanities Advanced Research Centre\\
Department of Classical Philology and Italian Studies\\
University of Bologna\\
Via Zamboni 32, 00126 Bologna, Italy\\
\texttt{aldo.gangemi@unibo.it}
\And
Silvio Peroni\\
Digital Humanities Advanced Research Centre\\
Department of Classical Philology and Italian Studies\\
University of Bologna\\
Via Zamboni 32, 00126 Bologna, Italy\\
\texttt{silvio.peroni@unibo.it}
\And
Francesco Poggi\\
Digital and Semantic Publishing Laboratory\\
Department of Computer Science and Engineering\\
University of Bologna\\
Mura Anteo Zamboni 7, 00126 Bologna, Italy\\
\texttt{francesco.poggi5@unibo.it}
\And
Valentina Presutti\\
Semantic Technologies Laboratory\\
Institute of Cognitive Sciences and Technologies\\
Italian National Research Council\\
Via San Martino della Battaglia 44, 00185 Rome, Italy\\
\texttt{valentina.presutti@cnr.it}
}

\begin{document}
\maketitle

\begin{abstract}
Alternative metrics (aka altmetrics) are gaining increasing interest in the scientometrics community as they can capture both the volume and quality of attention that a research work receives online. Nevertheless, there is limited knowledge about their effectiveness as a mean for measuring the impact of research if compared to traditional citation-based indicators. This work aims at rigorously investigating if any correlation exists among indicators, either traditional (i.e. citation count and $h$-index) or alternative (i.e. altmetrics) and which of them may be effective for evaluating scholars. The study is based on the analysis of real data coming from the National Scientific Qualification procedure held in Italy by committees of peers on behalf of the Italian Ministry of Education, Universities and Research.
\end{abstract}


\section{Introduction}
\label{sec:intro}
Altmetrics\footnote{\url{http://altmetrics.org/manifesto/}} have been introduced by~\cite{Priem2012} as the study and use of scholarly impact measures based on activity in online tools and environments. The term has also been used to describe the metrics themselves. According to~\cite{Priem2012} the study of altmetrics is a subset of both scientometrics and webometrics. Namely, this subset focuses more narrowly on scholarly influence as measured in online tools and environments, rather than on the Web more generally. The sources used for altmetrics are heterogeneous and include mentions and citations in blogs, Wikipedia, Twitter, Facebook or reader counts on social reference managers and bookmarking platforms. Altmetrics probably capture a broad research audience and can be used for measuring different flavours of impact. For example, according to~\cite{BeaverR78}, non-publishing readers (i.e. readers that do not publish research works) are estimated to constitute one third of the scientific community. Non-citation-based metrics are not novel. In fact, a lot of research has been carried out so far in order to understand the correlations between traditional citations and emerging metrics. Examples are ~\cite{Thelwall2008} for online presentations, \cite{Kousha2008} for online syllabi, \cite{Kousha2006,Meho2007} for Google Scholar citations, \cite{Kousha2009} for Google Book citations, and~\cite{Brody2006,Pinkowitz2002} for article downloads. 
Nevertheless, there is limited or none scientific evidence that altmetrics are valid proxies of either impact or utility although a few case studies have reported medium correlations between specific altmetrics and citation rates for individual journals or fields. In this work we want to investigate the following research questions:
\begin{itemize}
\item {\em RQ1}: what are the indicators, either traditional or alternative identifies, that share a common intended meaning?
\item {\em RQ2}: what kind of indicators can be used to provide effective suggestions about research quality?
\end{itemize}
In order to address the aforementioned research questions we conduct a study by using the data from the first National Scientific Qualification (NSQ\footnote{The Italian acronym is ASN, which stands for Abilitazione Scientifica Nazionale.}) held in Italy in 2016. The NSQ is the procedure for obtaining the national scientific qualification for the roles of associate and full professor in Italian Universities. Different committees composed of peers evaluate and assess the candidates for their scientific qualification in each discipline, called academic recruitment field. The NSQ is managed by the Italian Ministry of Education, Universities, and Research (MIUR). 

We perform our experiments on a representative sample of the NSQ 2016 dataset, which includes 689 and 1,174 candidates for the full professor level and associate professor level, respectively. Additionally, our sample contains 43,184 and 41,442 articles for the full professor level and associate professor level, respectively.
For each publication we gather traditional bibliographic indicators as well as altmetrics in order to:
\begin{itemize}
\item investigate the correlation between indicators, either traditional or alternative, i.e. {\em RQ1};
\item investigate the behaviour of traditional as well as alternative indicators used as features for automatically assessing scholars, i.e. {\em RQ2}.
\end{itemize}
The rest of the paper is organised in the following way: Section~\ref{sec:background} surveys some related works; Section~\ref{sec:material} describes the materials used for the experiments; Section~\ref{sec:exp_setup} describes the experiments; Section~\ref{sec:results} presents the results and their discussion; finally Section~\ref{sec:conclusion} presents our conclusions.

\section{Related work}
\label{sec:background}
An increasing number of research works have been carried out in order to study the relation between conventional bibliometric indicators and altmetrics.
Similarly, a number of approaches have been presented in literature to try to automatically predict the results of evaluation procedures by using bibliometric indicators.

{\bf Correlation between traditional indicators and altmetrics.}
Works like~\cite{Thelwall2013,Li2012_1,Li2012_2,Bar2012} have investigated the correlation between altmetrics and traditional bibliometric indicators. In general, these studies have found moderate agreement (i.e. $\sim$0.6 with Spearman correlation coefficient) with specific sources of altmetrics, i.e. Mendeley and Twitter. According to~\cite{Thelwall2018}, the number of Mendeley readers tends to correlate better with synchronous citation counts after a few years. These results have been confirmed also by the meta-analysis conducted by~\cite{Bornmann2015alternative} that concluded that the correlation with traditional citations for micro-blogging  is negligible, for blog counts it is small, and for bookmark counts from online reference managers, it is medium to large.
Nevertheless, none of these studies analyses the correlation between altmetrics and traditional bibliometric indicators by also taking into account quality assessment procedures performed by peers. These procedures are typical of many academic evaluation systems in different countries across the world. 
Notably, works like~\cite{Wouters2015metric,Ravenscroft2017measuring,Bornmann2018altmetrics} perform this kind of analysis. More specifically, \cite{Wouters2015metric} correlate different metrics with the output of the Research Excellence Framework (REF) held in 2016 in the UK. The REF is the reference system for assessing the quality of research in UK higher education institutions. The analysis is performed with different traditional and alternative metrics and the outcomes available for different research areas converge towards limited or none correlation.
Similarly, \cite{Ravenscroft2017measuring} find very low or negative correlation coefficients between altmetrics provided by Altmetric.com and REF scores concerning societal impact published by British universities in use case studies. Finally, the aim of the analysis carried out by \cite{Bornmann2018altmetrics} is twofold. In fact, first the authors in \cite{Bornmann2018altmetrics} measure the correlation between citation counts and altmetrics by using Principal Component Analysis and Factor Analysis. Then, they test the relationship between the dimensions and quality of papers using regression analysis on post-publication peer-review system of F1000Prime assessments. The results of the first part show how the count of Mendeley readers and tweets are related to citation count, while the regression analysis shows that only Mendeley readers and citation count are significantly related to quality. None of the aforementioned works present a comprehensive analysis investigating not only the correlation between traditional indicators and altmetrics, but also the correlation among the altmetrics themselves. The latter perspective is relevant in order to understand how to eventually use altmetrics as well as traditional indicators to support peers in quality assessment procedures of research outcomes.

{\bf Automatic prediction of evaluation procedures based on bibliometric indicators.}
A few works have focused so far on the problem of using bibliometric indicators to predict the results of evaluation procedures performed by peer.
An example is the work by~\cite{Vieira2014definition}, that uses a model based on a set of bibliometric indicators for the prediction of the ranking of applicants to an academic position as produced by a committee of peers. The results show that a very small number of indicators may lead to a robust prediction of about 75\%.
Another example is the work by~\cite{Ibanez2011predicting} that uses bibliometric indicators to train a N\"{a}ive Bayes algorithm to predict the $h$-index of Spanish professors for time horizon of four years. The results by~\cite{Ibanez2011predicting} show that it is easier to predict the $h$-index of the one-year time horizon and that it is easier to predict the $h$-index of junior professors than senior professors. Finally, \cite{Jensen2009testing} investigate how to predict career promotions by using a binomial regression model trained on quantitative indicators, such as the $h$-index, the number of publications and citations, etc. The evaluation performed on 3659 CNRS researchers from all disciplines between 2005 and 2008 shows a precision of 0.48. To the best of our knowledge, none of the aforementioned works use altmetrics features for their prediction algorithms.

\section{Material and method}
\label{sec:material}
In this section we present the data and how we process them in order to setup our experiment (cf. Section~\ref{sec:exp_setup}). More in detail, we explain how we constructed a knowledge graph from those data. Such a knowledge graph is modelled according to an ontology designed for modelling the relevant parts of the scholarly domain associated with research works, scholars, and indicators.

\subsection{Data from the Italian National Scientific Qualification}
\label{sec:data}
The National Scientific Qualification is a procedure for the university professor position recruiting that is held in Italy. It is based on the evaluation of the curricula submitted by candidate scholars, that are publicly published on the specific site associated with the procedure\footnote{ 
\url{http://abilitazione.miur.it/public/pubblicacandidati.php}}. 
The evaluation is performed by a national commission composed of peers that evaluates and assesses the candidates scientific qualification according to rigorous scientific criteria.

\begin{figure}[!ht]
\centering
\subfigure[Full Professor.]{\includegraphics[scale=0.4]{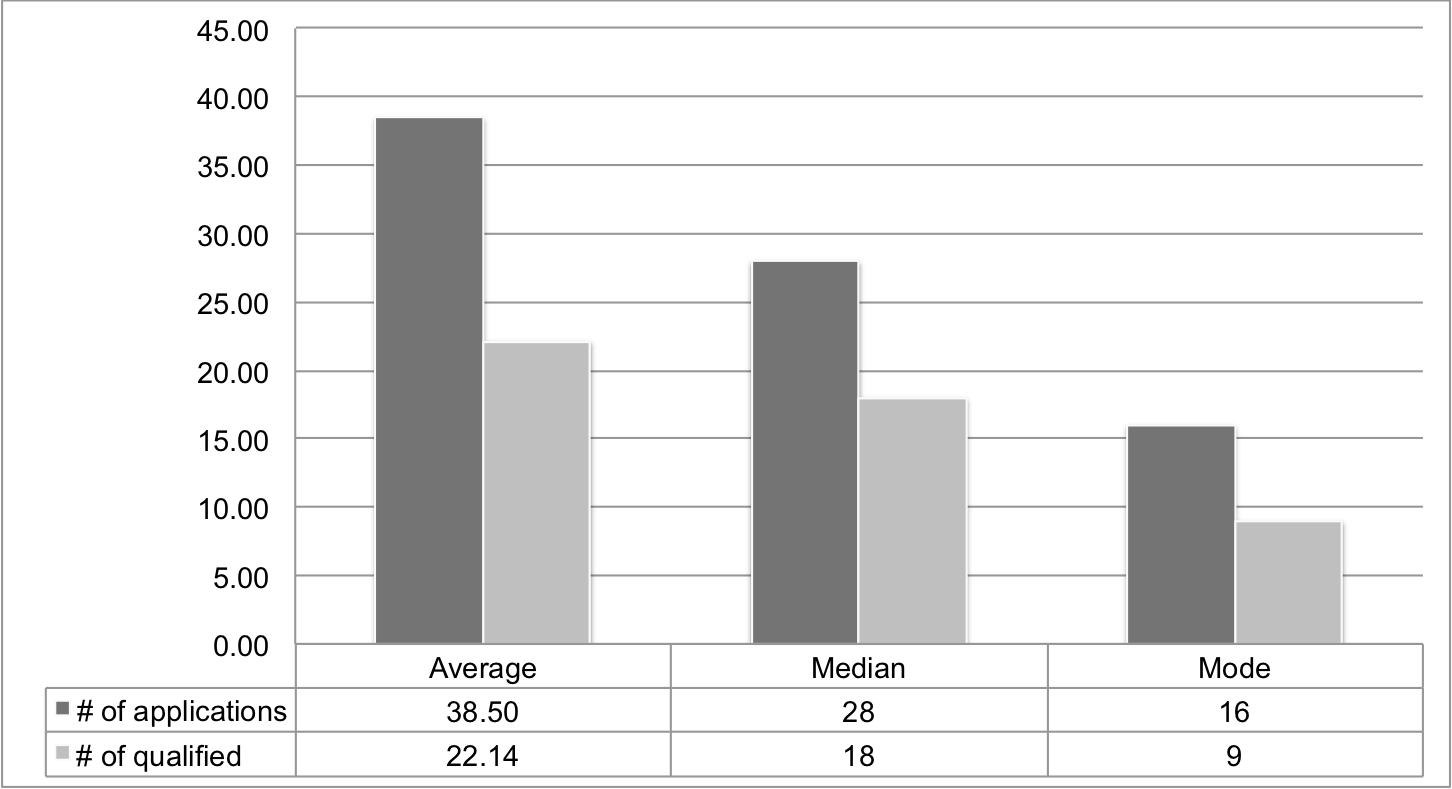}}\label{fig:applications_data_l1}
\subfigure[Associate Professor.]{\includegraphics[scale=0.4]{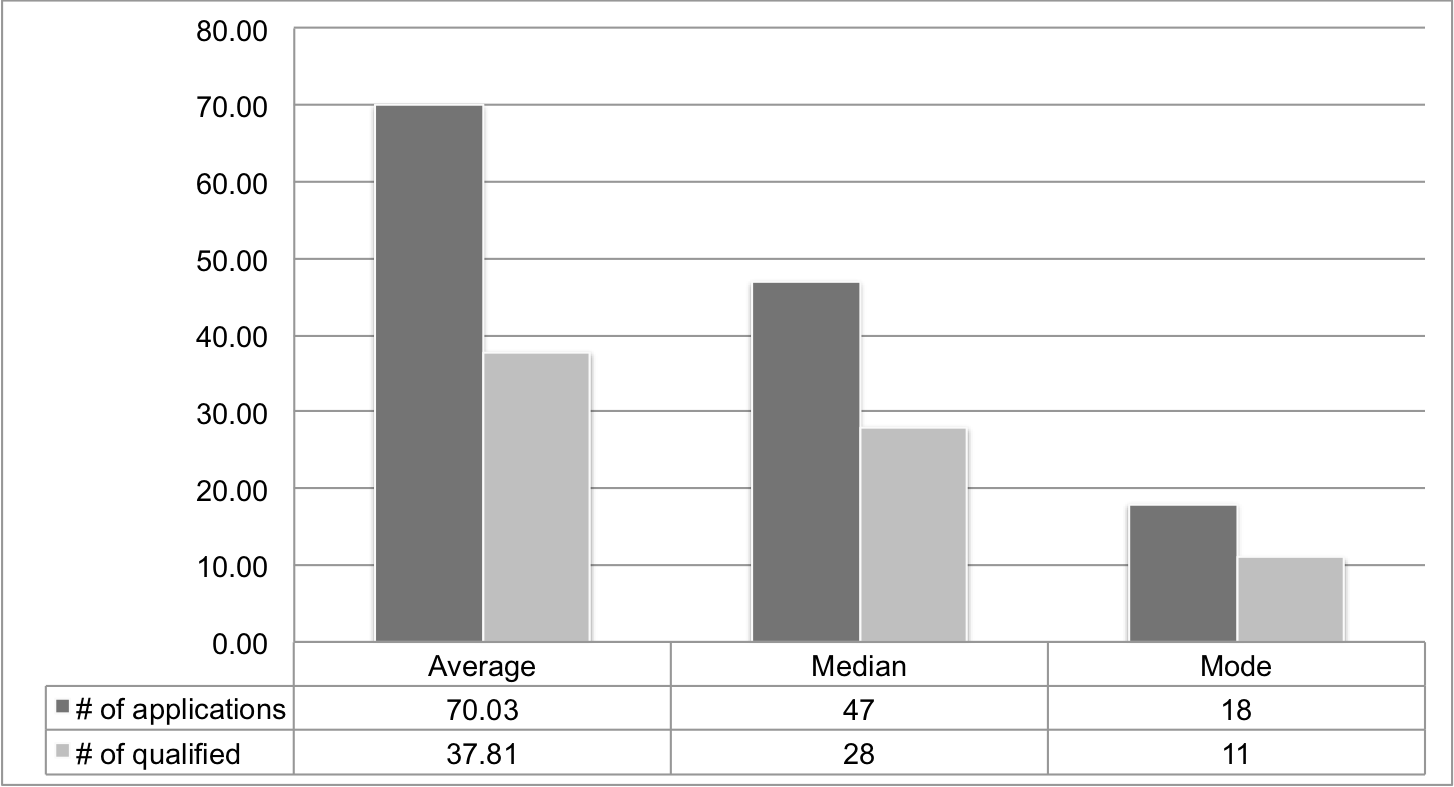}}\label{fig:applications_data_l2}
\caption{Aggregate data about the number of applications and scholars that obtained the qualification after the NSQ for the levels of Full Professor and Associate Professor, respectively.} \label{fig:applications_data}
\end{figure}

We collected the curricula related to the applications submitted to the first NSQ session for 2016. Then, we used the data extracted from the curricula submitted to both academic levels, i.e. full professor and associate professor, respectively. 
The available curricula cover 187 academic recruitment fields, according to the classification system formalised by the Italian Ministry of Education, Universities and Research (MIUR), which coordinates the NSQ. Figures~\ref{fig:applications_data_l1} and~\ref{fig:applications_data_l2} show aggregate data for both academic levels about the number of applications and scholars that obtained the qualification after the evaluation process of the NSQ. Those data report the average, the median and the mode for each perspective and academic level. We show aggregate data for reasons of readability. However, the interested reader can found detailed data reported for each academic recruitment field of the NSQ on-line\footnote{\url{https://figshare.com/articles/Input_data_of_MIRA/6854396}\label{fn:input-data}, doi: 10.6084/m9.figshare.6854396.v}. The original dataset counts 7,200 applications to the full professor level and 13,096 to the associate one. Thereby, each of the 187 distinct academic recruitment fields counts on average 38.5 applications and 70.03 applications to the full professor level and to the associate professor level, respectively.

\begin{figure}[!ht]
\centering
\subfigure[Number of publications.]{\includegraphics[scale=0.4]{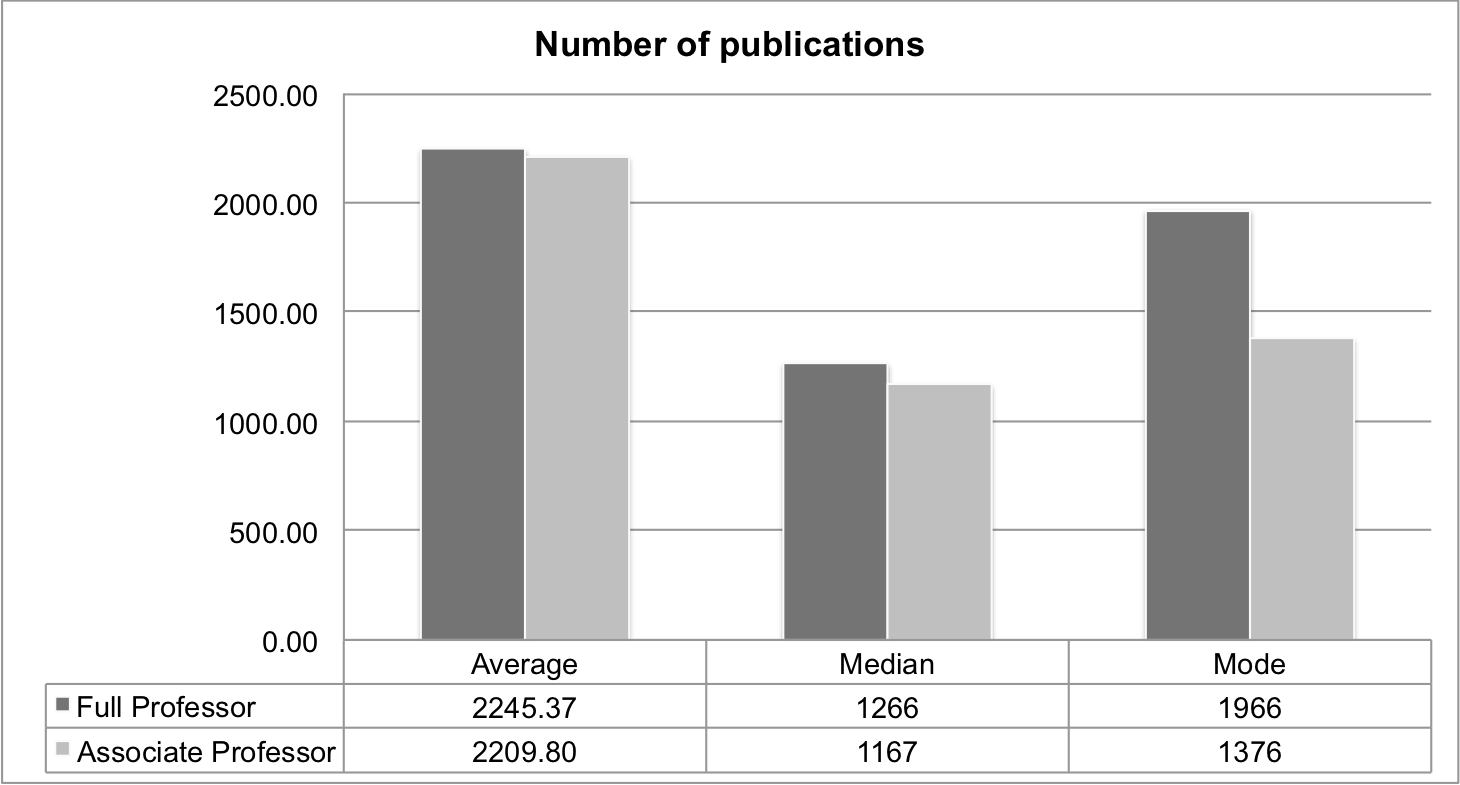}}\label{fig:applications_data_publications_global}
\subfigure[Number of publications per application.]{\includegraphics[scale=0.4]{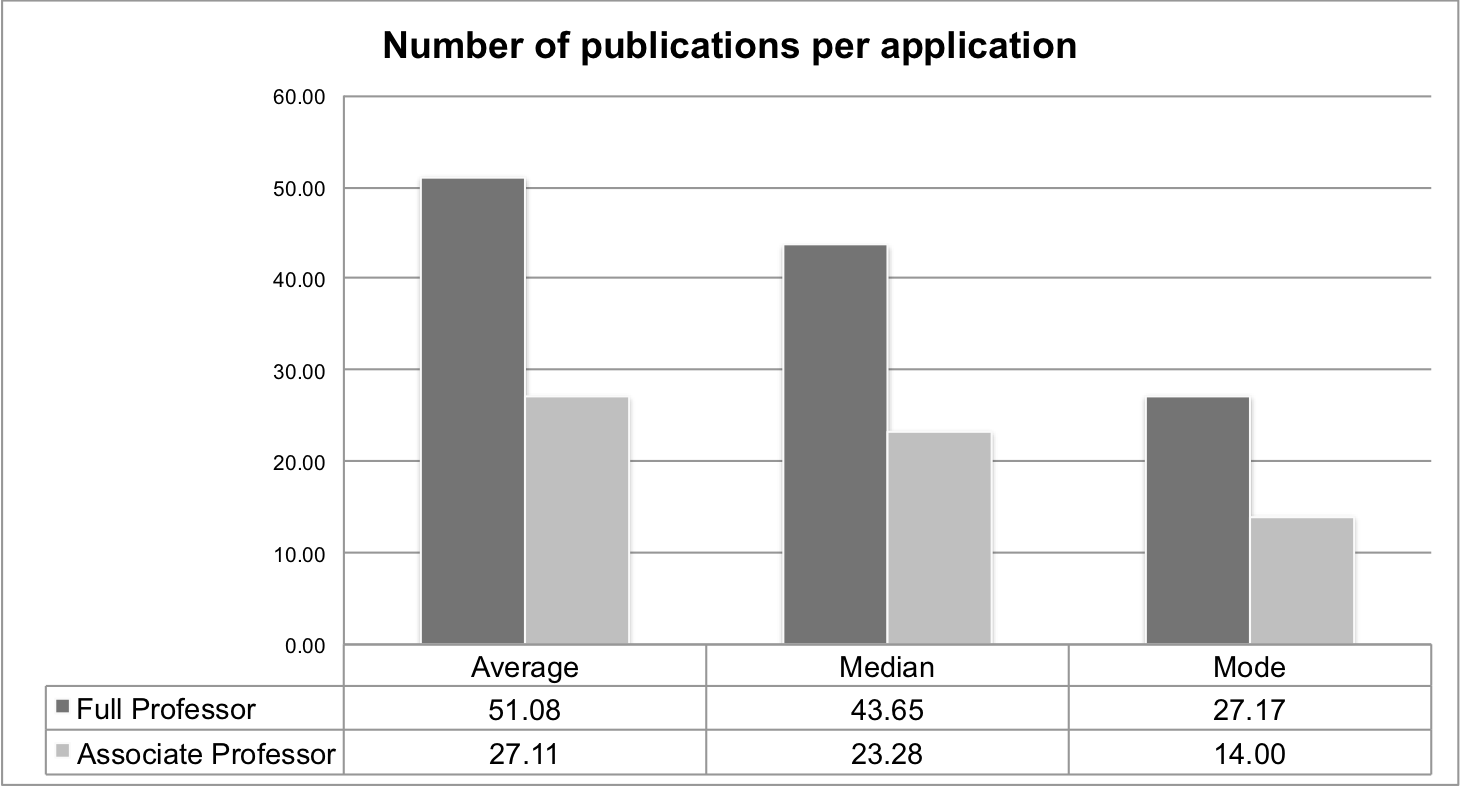}}\label{fig:applications_data_publications_app}
\caption{Aggregate data about the number of publications available from the applications and the perspective on publications available per application} \label{fig:applications_data_publications}
\end{figure}

The total number of publications gathered from the curricula is 419,884 and  413,232 for the two levels, respectively. Again, this means on average 2,245.37 and 2,209.8 publications for the two levels, respectively. Figure~\ref{fig:applications_data_publications_global} shows aggregate data about the the average number, the median, and the mode of publications recorded on the dataset globally. Instead, the average number of publication per curricula is 51.08 and 27.11 for the full professor and the associate professor levels, respectively. Figure~\ref{fig:applications_data_publications_app} shows these statistics along with the median and the mode. The interested reader can found the whole statistics aggregated by academic recruitment field on Figshare\footref{fn:input-data}.

\subsection{Data processing workflow}
\label{sec:workflow}
\begin{figure}[ht!] \centering
\includegraphics[scale=0.25]{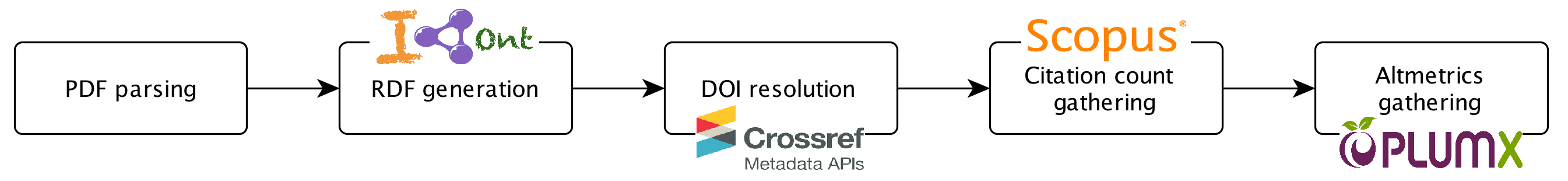}
	\caption{An overview on workflow that allows us to gather structured knowledge from the list of publications available from the PDF curricula.}
	\label{fig:workflow}
\end{figure}

The curricula that we used as data source are PDF files formatted according to the standard template provided by the NSQ procedure. This template is organised in three main sections, that is (i) the short list of the selected publications that are used by the evaluation committee of peers to assess the scientific contribution of a candidate in the specific academic recruitment field; (ii) the full list of publications that are used for computing indicators on Scopus or WoS, i.e. citations count and $h$-index; (iii) the full list of academic and professional titles that further characterise the candidate's academic profile for the peer evaluation. Among these sections we are interested in the full list of publications only. In fact, such list conveys all the relevant information useful for gathering structured knowledge about the publications and their corresponding indicators associated with a candidate. However, extracting structured knowledge from the list of publications available in the input PDF files is not straightforward. Accordingly, we designed a workflow to accomplish this task. Figure~\ref{fig:workflow} shows a graphical representation of the workflow we used.

The first step of the workflow is the PDF parsing, which allows us to identify the list of publications within a PDF and to pre-process such a list into a mediator data structure. This mediator data structure allows to extract the core metadata from a bibliographic record, i.e. the authors, the title, the publication venue (e.g. journal name or conference name), and, whenever possible, the digital object identifier (DOI). 

The structured representation of these core metadata is preparatory to the next step, which is the conversion to RDF of the bibliographic records. The resulting RDF is modelled by using the {\em Indicators Ontology} (I-Ont) and feeds a knowledge graph that will provide us with the basic building blocks for carrying out our analysis. I-Ont~\cite{NuzzoleseSaveSd} is an ontology for representing scholarly artefacts and their associated indicators, e.g. citation count or altmetrics such as the number of readers on Mendeley. I-Ont is designed as an OWL ontology and was originally meant for representing indicators associated with the papers available on ScholarlyData, which is the reference linked open dataset of the Semantic Web community about papers, people, organisations, and events related to its academic conferences. 

Once the data are available in the knowledge graph and modelled according to the I-Ont, the workflow requires a number of steps for enriching such a knowledge graph with DOIs, citation counts, and altmetrics. The DOIs are gathered by querying Crossref by means of its public API\footnote{\url{https://github.com/CrossRef/rest-api-doc}}. We query Crossref only for those articles which are not associated with a DOI in the knowledge graph as having articles associated with DOIs is mandatory for next steps. In fact, we use the DOIs as key parameters for gathering citation counts and atlmetrics by querying the REST API of Scopus\footnote{\url{https://dev.elsevier.com/tecdoc_cited_by_in_scopus.html}} and Plum Analytics\footnote{\url{https://github.com/PlumAnalytics/plum-api}} (PlumX), respectively. We use Scopus as it is the reference service used by the committees of peers for gathering candidates' indicators during the NSQ procedure. 

Instead, we use PlumX among the variety of altmetrics providers (e.g. Altmetric.com or ImpactStory) as, according to~\cite{Peters2014}, it is the service that registers the most metrics for the most platforms. Additionally PlumX is currently the service that covers the highest number of research work ($\sim$52.6M\footnote{Data retrieved from \url{https://plumanalytics.com/learn/about-metrics/coverage/} on June 2018.}) if compared to Altmetric.com ($>$5M\footnote{Data retrieved from \url{https://figshare.com/articles/Altmetric_the_story_so_far/2812843/1} on June 2018.}) and ImpactStory ($\sim$1M\footnote{Data retrieved from \url{https://twitter.com/Impactstory/status/731258457618157568} on June 2018.}). PlumX provides three different levels of analytics consisting of (i) the category, which provides a global view across different indicators that are similar in semantics (e.g. the mentions of a research work on social media); (ii) the metric, which identifies the indicator (e.g. the number of tweets); (iii) and the source, that basically allows to track the provenance of an indicator (e.g. the number of tweets on Twitter). From now on in the paper we refer to these levels as to the category-metric-source hierarchy. The categories are (i) the {\em Usage}, which represents a signal if anyone is reading an article or otherwise using a research; (ii) the {\em Captures}, which indicate that someone wants to come back to the work; (iii) the {\em Mentions}, which measure activities such as news articles or blog posts about research; (iv) the {\em Social Media}, which include tweets, Facebook likes, etc. that reference a research work; and (v) {\em Citations} contains both traditional citation indexes such as Scopus, as well as citations that help indicate societal impact such as Clinical or Policy Citations. A detailed description of the category-metric-source hierarchy can be found on the documentation provided by PlumX\footnote{\url{https://plumanalytics.com/learn/about-metrics/}}.

The workflow is implemented as a Java project and its source code is publicly available on GitHub\footnote{\url{https://github.com/anuzzolese/mira}}. The implementation of the workflow relies on the Apache PDFBox library~\footnote{\url{https://pdfbox.apache.org/}} to deal with PDF and on RDF4J~\footnote{\url{http://rdf4j.org/}} to produce and manipulate RDF.

\section{Evaluation}
\label{sec:exp_setup}
The aim of the experiment we present is twofold. Namely, we want to (i) analyse the correlation among indicators, either traditional or alternative and (ii) investigate the behavior of both traditional and alternative indicators used as features of a classification algorithm for automatically predicting the results of the NSQ by means of machine learning. 
The correlation analysis is designed to address {\em RQ1}, while the classification experiment is designed to address {\em RQ2}.

\subsection{Experimental setup}
\label{sec:experiment-datasample}
In order to carry on our experiment (i) we selected a representative subset of academic recruitment fields out of the 187 fields available from our input data (cf. Section~\ref{sec:data}); (ii) we applied the workflow described in Section~\ref{sec:workflow} for generating a knowledge graph; and (iii) we used the knowledge graph as the input for performing the correlation analysis and the automatic classification task.

The sample is composed of a representative subset of the academic recruitment fields out of the whole set of 187 of the original dataset. To perform a meaningful selection we applied the following rationale. First, we used the macro classification of the fields proposed by~\cite{Costas2015} as a reference model. Such a classification consists of the following macro-fields:
\begin{itemize}
\item Biomedical and health sciences;
\item Life and earth sciences;
\item Mathematics and computer science;
\item Natural sciences and engineering;
\item Social sciences and humanities.
\end{itemize}
\begin{center}
\begin{table}[!ht]
\caption{Mapping of recruitment academic fields to macro fields according as defined by~\cite{Costas2015}.} 
\label{tab:field-classification}
\resizebox{\textwidth}{!}{ 
\begin{tabular}{p{5cm}p{8cm}}
\centering
{\bf Macro field } & {\bf Recruitment academic fields } \\ \hline \hline
Biomedical and health sciences & 05-A1, 05-A2, 05-B1, 05-B2, 05-C1, 05-D1, 05-E1, 05-E2, 05-E3, 05-F1, 05-G1, 05-H1, 05-H2, 05-I1, 05-I2, 06-A1, 06-A2, 06-A3, 06-A4, 06-B1, 06-C1, 06-D1, 06-D2, 06-D3, 06-D4, 06-D5, 06-D6, 06-E1, 06-E2, 06-E3, 06-F1, 06-F2, 06-F3, 06-F4, 06-G1, 06-H1, 06-I1, 06-L1, 06-M1, 06-M2, 06-N1, 06-N2 \\ \hline
Life and earth sciences & 04-A1, 04-A2, 04-A3, 04-A4, 07-A1, 07-B1, 07-B2, 07-C1, 07-D1, 07-E1, 07-F1, 07-G1, 07-H1, 07-H2, 07-H3, 07-H4, 07-H5, 07-I1 \\ \hline
Mathematics and computer science & 01-A1, 01-A2, 01-A3, 01-A4, 01-A5, 01-A6, 01-B1 \\ \hline
Natural sciences and engineering & 02-A1, 02-A2, 02-B1, 02-B2, 02-C1, 02-D1, 03-A1, 03-A2, 03-B1, 03-B2, 03-C1, 03-C2, 03-D1, 03-D2, 08-A1, 08-A2, 08-A3, 08-A4, 08-B1, 08-B2, 08-B3, 08-C1, 08-D1, 08-E1, 08-E2, 08-F1, 09-A1, 09-A2, 09-A3, 09-B1, 09-B2, 09-B3, 09-C1, 09-C2, 09-D1, 09-D2, 09-D3, 09-E1, 09-E2, 09-E3, 09-E4, 09-F1, 09-F2, 09-G1, 09-G2, 09-H1 \\ \hline
Social sciences and humanities & 10-A1, 10-B1, 10-C1, 10-D1, 10-D2, 10-D3, 10-D4, 10-E1, 10-F1, 10-F2, 10-F3, 10-F4, 10-G1, 10-H1, 10-I1, 10-L1, 10-M1, 11-A1, 11-A2, 11-A3, 11-A4, 11-A5, 11-B1, 11-C1, 11-C2, 11-C3, 11-C4, 11-C5, 11-D1, 11-D2, 11-E1, 11-E2, 11-E3, 11-E4, 12-A1, 12-B1, 12-B2, 12-C1, 12-C2, 12-D1, 12-D2, 12-E1, 12-E2, 12-E3, 12-E4, 12-F1, 12-G1, 12-G2, 12-H1, 12-H2, 12-H3, 13-A1, 13-A2, 13-A3, 13-A4, 13-A5, 13-B1, 13-B2, 13-B3, 13-B4, 13-B5, 13-C1, 13-D1, 13-D2, 13-D3, 13-D4, 14-A1, 14-A2, 14-B1, 14-B2, 14-C1, 14-C2, 14-C3, 14-D1 \\ \hline
\end{tabular}
}
\end{table}
\end{center}  
Hence, we mapped each academic recruitment field to one of these macro-field. Table~\ref{tab:field-classification} reports the result of such a mapping. The human readable labels corresponding to the codes of the recruitment fields reported in Table~\ref{tab:field-classification} is available on the Figshare repository~\footnote{\url{https://figshare.com/articles/Academic_Recruitment_Fields/6860576}, doi: 10.6084/m9.figshare.6860576.v1} associated with this work. Then, we associated each academic recruitment field with:
\begin{itemize}
\item the number of applications $A$ to the field $i$, i.e. $A_{i}$;
\item the total number of publications $P$ available for a field $i$, i.e. $P_{i}$. This number is obtained by counting all the publications available from the curricula submitted to a field.
\end{itemize}
$A_{i}$ and $P_{i}$ are computed for both the first (i.e. full professor) and the second (i.e. associate professor) academic levels. Hence, we use $A_{i,1}$ and $P_{i,1}$, and $A_{i,2}$ and $P_{i,2}$ to identify the number of applications and publications with respect to the first and the second level of the $i$-th academic recruitment field, respectively.
Then, we computed the variance $\sigma^{2}$ for $A_{i,1}$, $A_{i,2}$, $P_{i,1}$, and $P_{i,2}$, i.e. $\sigma^{2}_{A_{i,1}}$, $\sigma^{2}_{A_{i,2}}$, $\sigma^{2}_{P_{i,1}}$, and $\sigma^{2}_{P_{i,2}}$. We remind that the variance allows to measure how far a certain variable in a set is spread out from the average value computed across all the values in the same set. $AP_{i}$ measures the mean over the four variances as shown in Equation~\ref{eq:ap}.

\begin{equation}
\label{eq:ap}
AP_{i} = \frac{\sigma^{2}_{A_{i,1}} + \sigma^{2}_{P_{i,1}} + \sigma^{2}_{A_{i,2}} + \sigma^{2}_{P_{i,2}}}{4}
\end{equation}
Intuitively, $AP_{i}$ allows us to measure how far the numbers of submissions and publications in a field are spread out from their average value computed across all the fields. Basically, $AP_{i}$ is a tool for identifying a candidate academic recruitment field for each macro-field. This candidate fields are then used for carrying out the experiment. Accordingly, for each macro-field, we selected the candidate academic recruitment field with the maximum value of $AP_{i}$. This maximum value was obtained by applying Equation~\ref{eq:apmax}, namely:
\begin{equation}
\label{eq:apmax}
AP_{max} = \max \{AP_{1}, AP_{2}, ..., AP_{n}\}
\end{equation}

%

Accordingly, we used $AP_{max}$ in order to identify the academic recruitment field with the maximum $AP$ value, which are: 
\begin{itemize}
\item 01-B1, i.e. computer science, for the macro-field mathematics and computer science with $AP_{max}=0.56$;
\item 04-A1, i.e. geochemistry, mineralogy, petrology, vulcanology, geo-resource and applications, for the macro-field life and earth sciences with $AP_{max}=0.34$;
\item 06-N1, i.e. technology and methodology in medicine and nursing sciences, for the macro-field biomedical and health sciences with $AP_{max}=0.45$;
\item 09-H1, i.e. information processing systems, for the macro-field natural sciences and engineering with $AP_{max}=0.58$;
\item 13-A1, i.e. political economy, for the macro-field social sciences and humanities with $AP_{max}=0.51$.
\end{itemize}
It is worth noticing that the academic field 11-E1 (i.e. general psychology, psychobiology and psychometrics) is the one with the highest $AP_{max}$ score (i.e. 0.56) within the  macro field ``Social sciences and humanities''. Nevertheless, we selected the field 13-A1 (i.e. political economy) instead of 11-E1 as we wanted to include a non-bibliometric field in the sample. As a matter of fact 13-A1 is the non-bibliometric field with the highest $AP_{max}$ value recorded (0.51). The rationale behind this choice is straightforward, namely, we want to assess to what extent bibliometric indicators (i.e. citation count and $h$-index) along with altmetrics can be used to assess the research impact in academic recruitment fields that are non-bibliometric. 

Finally, we applied the workflow described in Section~\ref{sec:workflow} and the knowledge graph, i.e. MIRA-KG, resulting from this activity counts of 21,093,433 RDF triples. MIRA-KG is available on Figshare\footnote{\url{https://figshare.com/articles/MIRA_KG/6860240}, doi: 10.6084/m9.figshare.6860240.v1} as a graph database enabled by Blazegraph\footnote{\url{https://www.blazegraph.com/}}, an ultra-scalable, high-performance graph database.

\subsection{Experiments}
\label{sec:experiments}
We designed two experiments for evaluating indicators and addressing {\em RQ1} and {\em RQ2} (cf. Section~\ref{sec:intro}).

{\bf Correlation among indicators.}
We use the sample Pearson correlation coefficient (i.e. $r$) as the measure to assess the linear correlation between pairs of sets of indicators (i.e. {\em RQ1}). The Pearson correlation coefficient is widely used in literature and has a value ranging from +1 to -1, where, +1 indicates total positive linear correlation, 0 indicates no linear correlation, and -1 indicates total negative linear correlation. 

In order to compute $r$, we generated vectors for all the publications available in the sample dataset. The elements of a vector are the the indicators associated with the specific publication that the vector provides a representation of. We filled elements with 0 when the specific indicators corresponding to those elements were not available for a certain publication. The latter condition was necessary in order to have equal vectors in size.
We remark that PlumX provides a three-tiered hierarchy based on the chain category-metric-source (e.g. Captures-Reader-Mendeley). In such a hierarchy the category is associated with multiple metrics and a metric is associated with multiple sources. This led to three different conditions of our experiment. In fact, we built three different sets of vectors, that is one for the indicators summarised by the categories, one for indicators summarised the metrics, and one for the sources. In all sets of vectors we added the citation counts from Scopus. Accordingly, the correlation analysis ranges from a broader perspective, which includes PlumX's categories, to a narrower one, which includes PlumX's sources.
Once built the indicators vectors, we computed the $r$ coefficients for each academic level and for each academic recruitment field part of the sample. Then, we averaged those coefficients among the academic recruitment field by distinguishing between the full professor level and the associate professor level.

Finally, we defined another condition for our experiment. Namely, we computed the correlation coefficients at author level by using $h$-index scores as indicators. Accordingly, we computed the $h$-index scores for altmetrics with the conventional $h$-index function, but using altmetrics as citation count values. At the same time we computed traditional $h$-index scores by using the Scopus citation count available at article level as input.

{\bf Automatic assessment of scholars}
The second experiment we designed is based on machine learning. This experiment analyses to what extent bibliometric indicators can be used for training a classification algorithm (i.e. {\em RQ2}). The algorithm automatically predicts if a certain candidate can be qualified to the academic recruitment field and with specific regards to the academic level she applied to. The metrics we used are the Scopus citation count, the altmetrics from PlumX, and $h$-index scores computed with citation counts and altmetrics. We used Na\"{i}ve Bayes as classification algorithm and the results for the academic recruitment fields of our sample available from the first session of the NSQ 2016 as dataset for training and testing.

Na\"ive Bayes is a probabilistic technique that estimates the probability of a class being associated with a vector of words or \textit{posterior probability}. Given a vector and a set of classes, the class that maximises this probability is associated with the vector. Na\"ive Bayes assumes that the attributes are independent, namely, that the probability of an instance having a certain value, given a class does not depend on the values assumed by the other attributes.
For each academic recruitment field and academic level (i.e. Full Professor and Associate Professor) we have two nominal classes to use for the classification algorithm that come from the NSQ procedure, i.e. ``Qualified'' or ``Not Qualified''. Table~\ref{tab:classification-dataset} report the details about the dataset used for experimenting with the Na\"{i}ve Bayes classifier. Thus, we evaluate 10 classifiers (i.e. a classifier for each academic level available for each academic recruitment field) by performing 10 runs of a 10-fold cross validation. The performance of the classifier is obtained by averaging precision, recall, and F-Measure resulting from the 10 runs.
\begin{center}
\begin{table}[!ht]
\centering
\caption{Dataset for binary classification.} 
\label{tab:classification-dataset}
\resizebox{\textwidth}{!}{ 
\begin{tabular}{p{1.5cm}||p{1.5cm}p{2.5cm}||p{1.5cm}p{2.5cm}}
\centering
 & \multicolumn{2}{c}{\bf Full Professor} & \multicolumn{2}{c}{\bf Associate Professor} \\ \hline\hline
 & {\bf Qualified } & {\bf Not Qualified } & {\bf Qualified } & {\bf Not Qualified } \\ \hline \hline
{\bf 01-B1 } & 75 & 160 & 94 & 213 \\ \hline
{\bf 04-A1 } & 37 & 14 & 63 & 34 \\ \hline
{\bf 06-N1 } & 43 & 94 & 111 & 265 \\ \hline
{\bf 09-H1 } & 63 & 108 & 113 & 158 \\ \hline
{\bf 13-A1 } & 49 & 46 & 50 & 73 \\
\end{tabular}
}
\end{table}
\end{center}  

\section{Results and discussion}
\label{sec:results}
Section~\ref{sec:results-corr} reports the results obtained for the correlation analysis. Instead, Section~\ref{sec:results-class} reports the results obtained from the classification experiment.

\subsection{Correlation}
\label{sec:results-corr}
The correlation analysis takes into account separately the results obtained for the full professor academic level and those obtained for the associate professor academic level. We remind that the indicators are those associated with the research works gathered from the curricula submitted to the first session of the Italian NSQ procedure held in 2016. More specifically, these indicators are limited to the five academic recruitment fields that identify our experimental sample as described in Section~\ref{sec:experiment-datasample}.

\begin{figure}[ht!] \centering
\subfigure[Full professor.]{\includegraphics[width=0.49\textwidth]{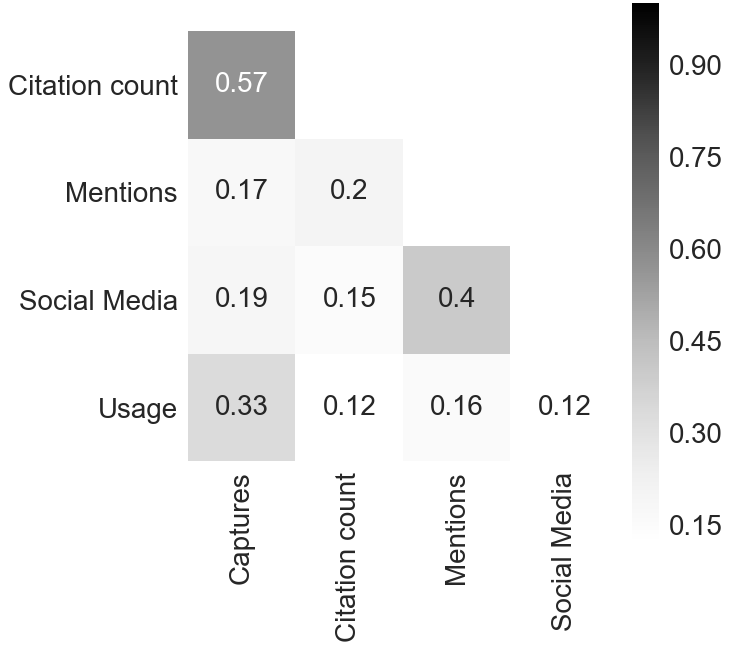}\label{fig:corr-cat-level-1}}
\subfigure[Associate professor.]{\includegraphics[width=0.49\textwidth]{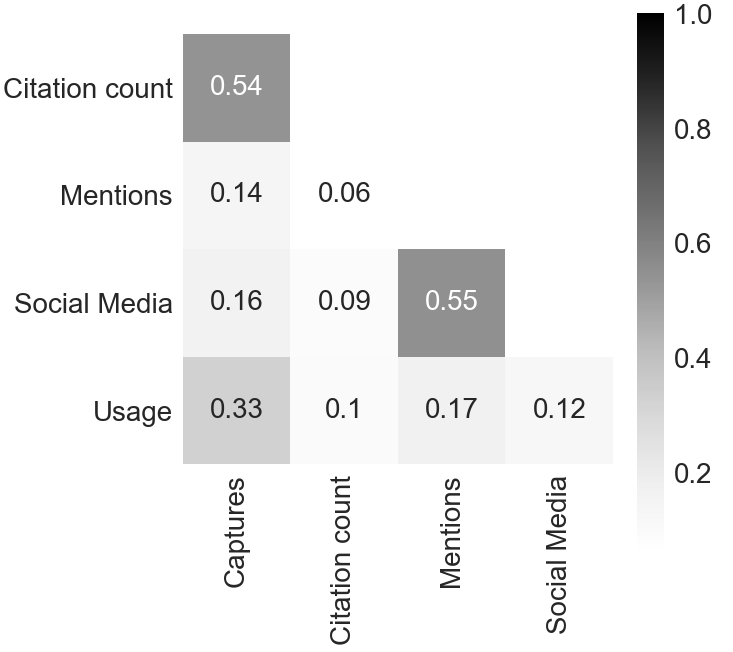}\label{fig:corr-cat-level-2}}
\caption{Correlation coefficients among category-level indicators gathered for both academic levels.}
\label{fig:corr-cat-levels}
\end{figure}

{\bf Category perspective.} Figures~\ref{fig:corr-cat-level-1} and~\ref{fig:corr-cat-level-2} show two confusion matrixes with the results obtained by computing the mean sample Pearson correlation coefficient (i.e. $r$) among all the indicators available in the categories as provided by PlumX. The $r$ coefficients reported in the confusion matrixes are obtained by means of pairwise comparisons among the categories of indicators as described in Section~\ref{sec:experiments}. According to the results, the highest correlation coefficients are those obtained between citation counts and captures, and between social media and mentions. In fact, for the full professor level, we recorded $r=0.57$, with statistical significance $p<0.01$ and standard error $SE_{r}=\pm 0.01$, and $r=0.4$, with $p<0.01$ and $SE_{r}=\pm 0.02$, as the correlations coefficients between citation counts and captures, and between social media and mentions, respectively. Similarly, for the associate professor level, we recorded $r=0.54$, with $p<0.01$ and $SE_{r}=\pm 0.01$, and $r=0.55$, with $p<0.01$ and $SE_{r}=\pm 0.01$, as the correlations coefficients between citation counts and captures, and between social media and mentions, respectively. The $p$-values for measuring the statistical significance are computed by using the Student's $t$-distribution. 
The whole result sets for the full and associate professor levels including the $r$ coefficients, $p$-values, and standard error computed for each academic recruitment filed part of the sample are published as spreadsheets on Figshare\footnote{The result set for the full professor level is available at \url{https://figshare.com/articles/Correlations_of_indicator_categories_for_Full_Professor/6882467}, doi: 10.6084/m9.figshare.6882467.v1. The result set for the associate professor level is available at \url{https://figshare.com/articles/Correlations_of_indicator_categories_for_Associate_Professor/6882488}, doi: 10.6084/m9.figshare.6882488.v1.}.

{\bf Metric perspective.} As we recorded the highest correlation coefficients for the pairs ``citation count-captures'' and ``social media-mentions'' we go deeper in the analysis by reporting the correlation coefficients obtained for the metrics, namely the altmetrics gathered from different sources and grouped by what PlumX defines metrics (e.g. number of readers on on-line tools). The metrics available in our dataset for the captures category are (i) the readers, which record the number of people who have added a specific research work to their library/briefcase; (ii) and the export-saves, which record the number of times a citation has been exported direct to bibliographic management tools or as file downloads, and the number of times a citation/abstract and HTML full text (if available) have been saved, emailed or printed. Similarly, the the metrics available in our dataset for the social media category are (i) the +1s, which record the number of times a research work has gotten a +1 on platform like Google Plus; (ii) the shares, likes \& comments, which record the number of times a link was shared, liked or commented on on social platforms; and (iii) the tweets, which record the number of tweets and retweets that mention a specific research work. Finally, the metrics available in our dataset for the mentions category are (i) the blog mentions, which record the number of blog posts written about a research work; (ii) the comments, which record the number of comments made about a research work; the economic blog mentions, which record the number of blog posts written about a research work within the economics discipline; (iii) the links, which record the number of references found to a research work; (iv) the news mentions, which record the number of news articles written about a research work; and (v) the reviews, which record the number of reviews written about a research work.

\begin{figure}[ht!] \centering
\subfigure[Full professor.]{\includegraphics[width=0.49\textwidth]{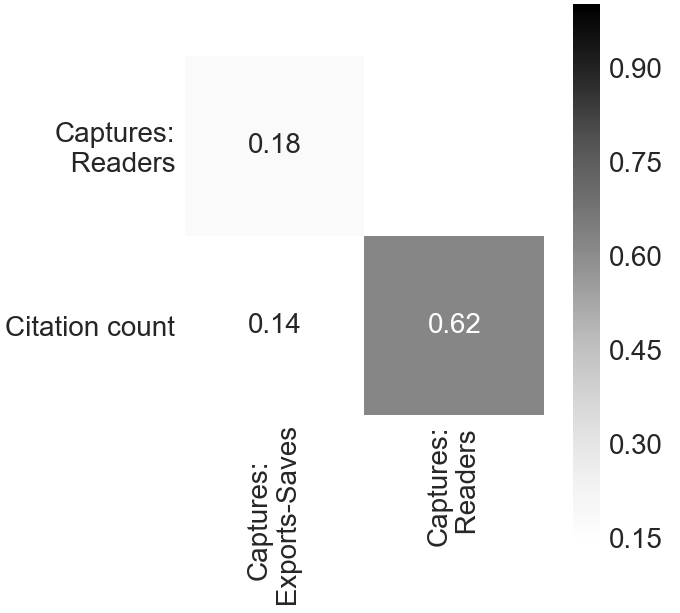}\label{fig:corr-captures-citations-level-1}}
\subfigure[Associate professor.]{\includegraphics[width=0.49\textwidth]{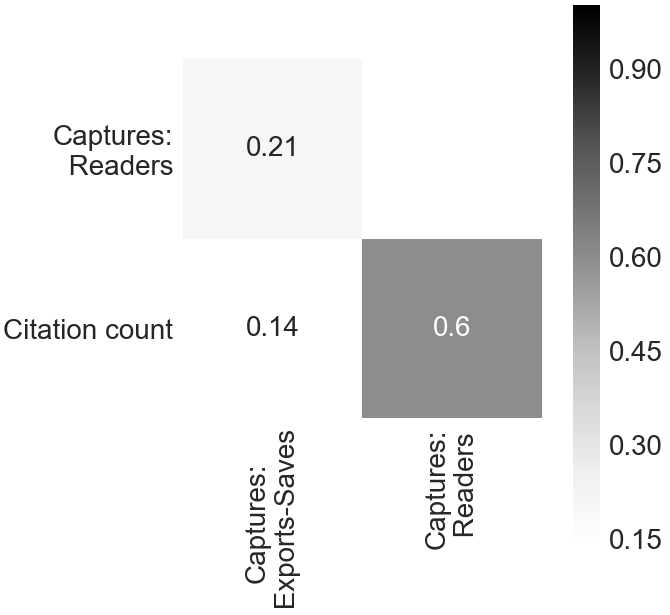}\label{fig:corr-captures-citations-level-2}}
\caption{Correlation coefficients between captures and citation counts gathered for both academic levels.}
\label{fig:cat-captures-citations-levels}
\end{figure}

Figure~\ref{fig:corr-captures-citations-level-1} and Figure~\ref{fig:corr-captures-citations-level-2} show the correlation coefficients computed among the metrics part of the pair ``citation count-captures'' for the academic level of full professor and associate professor, respectively. For these metrics we recorded the highest correlation coefficients between citation counts and readers. In fact, for the academic level of full professor the correlation is $r=0.62$, with $p<0.01$ and $SE_{r}=\pm 0.01$. Instead, for the academic level of associate professor we recorded $r=0.6$, with $p<0.01$ and $SE_{r}=\pm 0.01$. The full result set with correlation coefficients, $p$-values and standard errors is available of Figshare both for the full professor level\footnote{\url{https://figshare.com/articles/Correlations_among_citation_count_and_readers_for_Full_Professor/6882566}, doi: 10.6084/m9.figshare.6882566.v1} and the associate professor level\footnote{\url{https://figshare.com/articles/Correlations_among_citation_count_and_readers_for_Associate_Professor/7399295}, doi: 10.6084/m9.figshare.7399295.v1}.

\begin{figure}[ht!] \centering
\subfigure[Full professor.]{\includegraphics[width=0.49\textwidth]{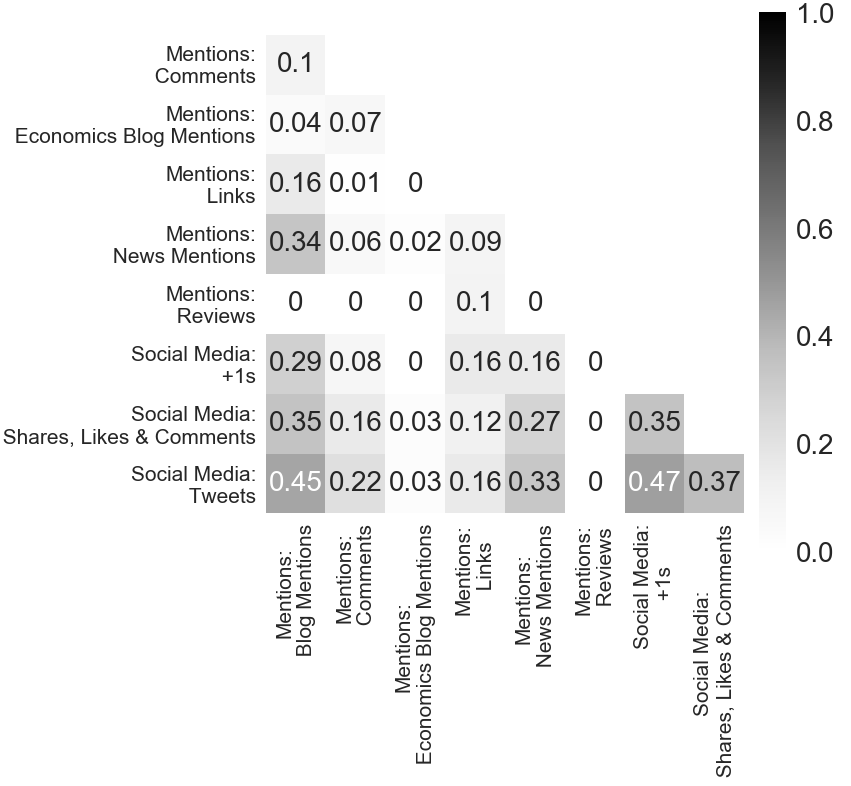}\label{fig:corr-mentions-social-media-level-1}}
\subfigure[Associate professor.]{\includegraphics[width=0.49\textwidth]{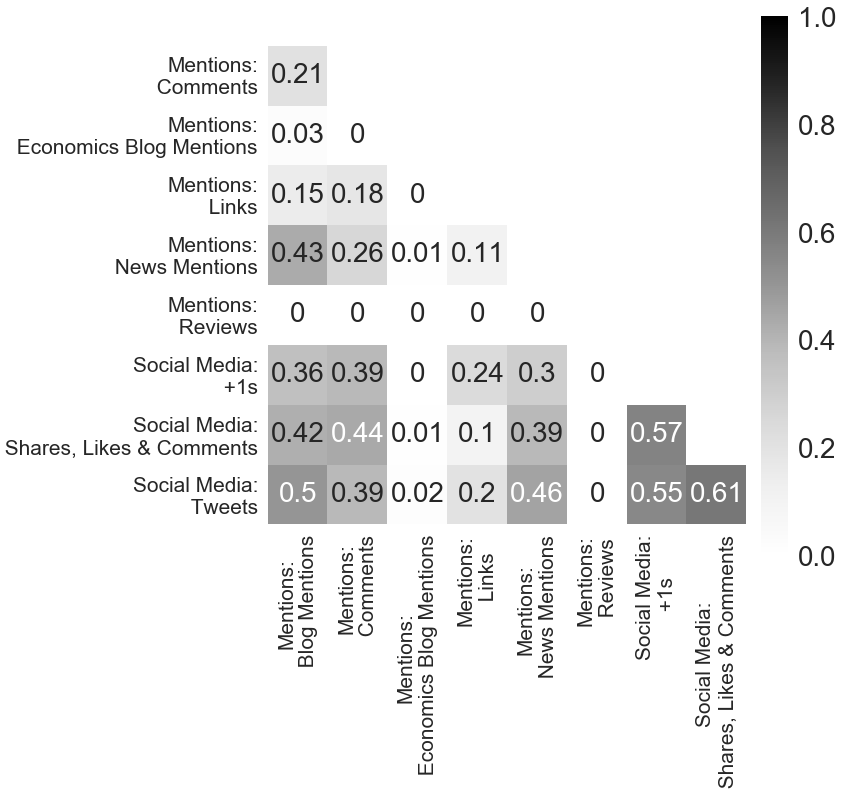}\label{fig:corr-mentions-social-media-level-2}}
\caption{Correlation coefficients between mentions and social media gathered for both academic levels.}
\label{fig:cat-mentions-social-media-levels}
\end{figure}

Figure~\ref{fig:corr-mentions-social-media-level-1} and Figure~\ref{fig:corr-mentions-social-media-level-2} show the correlation coefficients computed among the metrics, part of the pair ``social meida-mentions''. Again, those coefficients were computed both of full professor level and the associate one. In the analysis of the results we omit the correlation coefficients computed between metrics belonging to the same category, e.g. tweets and +1s. In fact, we assume that their correlation is not relevant for our investigation as they already represent indicators that are similar in kind. Accordingly, for the full professor level we recorded the highest correlation coefficient between tweets and blog mentions ($r=0.45$, $p<0.01$, and $SE_{r}=\pm 0.01$). We observe a similar scenario for the associate professor level. In fact, the highest correlation coefficient is recorded between tweets and blog mentions ($r=0.5$, $p<0.01$, and $SE_{r}=\pm 0.01$). Additionally, for the associate professor level we observe relatively high correlation coefficients between shares, likes \& comments and blog mentions ($r=0.42$, $p<0.01$, and $SE_{r}=\pm 0.01$) and between shares, likes \& comments ($r=0.44$, $p>0.1$, and $SE_{r}=\pm 0.01$).

\begin{figure}[ht!] \centering
\subfigure[Full professor.]{\includegraphics[width=0.45\textwidth]{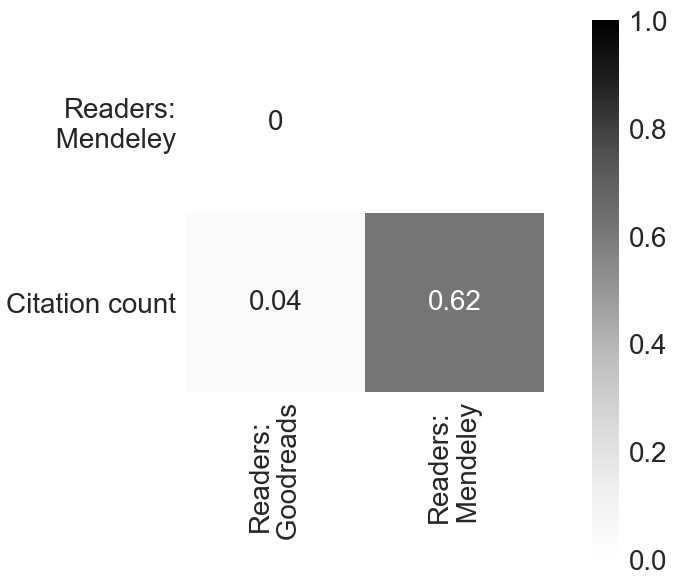}\label{fig:corr-readers-citations-level-1}}
\subfigure[Associate professor.]{\includegraphics[width=0.45\textwidth]{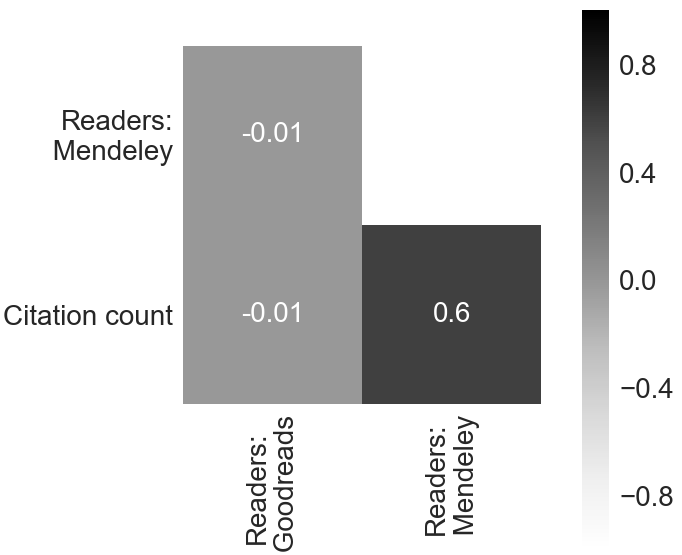}\label{fig:corr-readers-citations-level-2}}
\caption{Correlation coefficients between mentions and social media gathered for both academic levels.}
\label{fig:corr-readers-citations-levels}
\end{figure}

{\bf Source perspective.} If we investigate further the correlation by using original sources where altmetrics were gathered from, we observe good correlation coefficients computed between the citation counts and the readers fully depend on the number of reads tracked on Mendeley. 
In fact, Figure~\ref{fig:corr-readers-citations-level-1} shows that, for the full professor level, the correlation recorded between the citation counts and the number of readers on Mendeley ($r=0.62$, with $p<0.01$ and $SE_{r}=\pm 0.01$) is as high as that recorded for the whole readers metric (cf. Figure~\ref{fig:corr-captures-citations-level-1}). 

We observe the same scenario for the associate professor level. In fact, the correlation coefficient recorded between the citation counts and the number of readers on Mendeley is $r=0.6$, with $p<0.01$ and $SE_{r}=\pm 0.01$ (cf. Figure~\ref{fig:corr-readers-citations-level-2}). 
The detailed spreadsheets with all the results are published on Figshare both for the full professor level\footnote{\url{https://figshare.com/articles/Correlations_among_indicator_sources_for_Full_Professor/6882911}\label{fn:sources-level-1}, doi: 10.6084/m9.figshare.6882911.v1} and the associate professor level\footnote{\url{https://figshare.com/articles/Correlations_among_indicator_sources_for_Associate_Professor/6882917}\label{fn:sources-level-2}, doi: 10.6084/m9.figshare.6882917.v1}, respectively.
We do not present the results related to the sources associated with the social media and the mentions as each of their corresponding metric is based on a single source only, i.e. Twitter for tweets, Google Plus for +1s, Facebook for shares, likes \& comments, and a generally labelled source ``blog'' for blog mentions. Nevertheless, the interested readers might refer to the detailed spreadsheets published on Figshare both for the full professor level\footref{fn:sources-level-1} and the associate professor level\footref{fn:sources-level-2}, respectively.



{\bf Author perspective.} As described in Section~\ref{sec:experiments} we also derive $h$-index scores for the candidates to the first session of the Italian NSQ held in 2016. Those $h$-index scores are computed by using citation counts (i.e. traditional $h$-index) and altmetrcs based on PlumX categories. Accordingly, we have $h$-index scores for citation counts tracked by Scopus and for Capture, Mentions, Social Media, and Usage tracked by PlumX. 

\begin{figure}[ht!] \centering
\subfigure[Full professor.]{\includegraphics[width=0.45\textwidth]{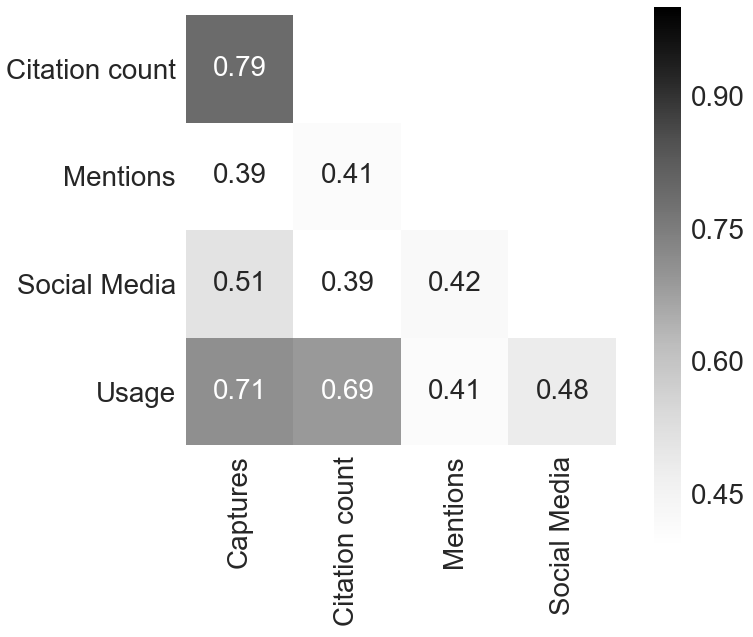}\label{fig:corr-hindex-level-1}}
\subfigure[Associate professor.]{\includegraphics[width=0.45\textwidth]{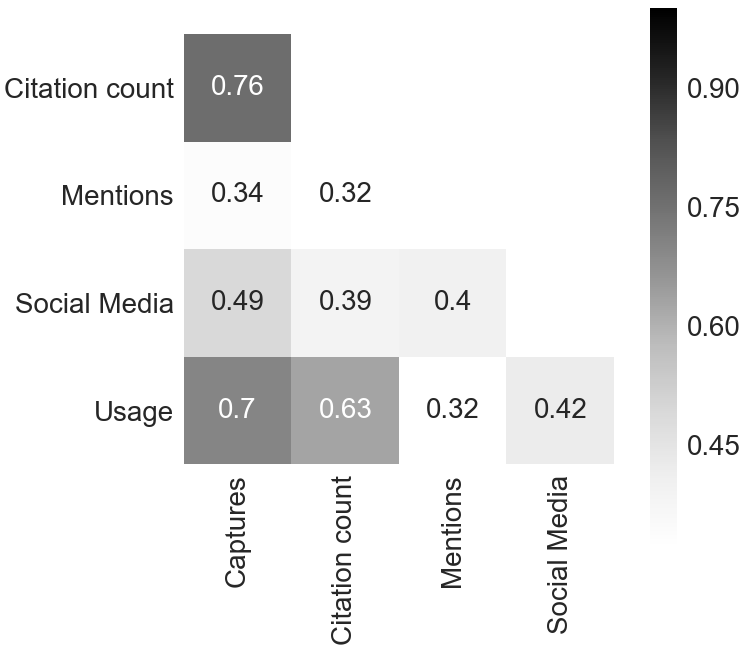}\label{fig:corr-hindex-level-2}}
\caption{Correlation coefficients among h-indexes computed among metrics for both academic levels.}
\label{fig:corr-hindex-levels}
\end{figure}

Figure~\ref{fig:corr-hindex-level-1} and Figure~\ref{fig:corr-hindex-level-2} show the correlation coefficients between the $h$-indexes based on the citation counts and on the altmetrics categories for the full professor level and the associate professor level, repsectively. According to the results, the author level indicators correlate with higher coefficients. In fact, with regards to the full professor level, we recorded high correlation between (i) the citation counts and the captures ($r=0.79$, $p<0.01$, and $SE_{r}=\pm 0.06$) and (ii) the usage and the capture ($r=0.71$, $p<0.01$, and $SE_{r}=\pm 0.06$). Similarly, with regards to the associate professor level, we recorded high correlation between (i) the citation counts and the captures ($r=0.76$, $p<0.01$, and $SE_{r}=\pm 0.04$); (ii) the usage and the capture ($r=0.7$, $p<0.01$, and $SE_{r}=\pm 0.05$); and the citation counts and the usage ($r=0.63$, $p<0.01$, and $SE_{r}=\pm 0.05$).



\subsection{Automatic classification}
\label{sec:results-class}
The analysis of the correlation coefficients investigates similarities among indicators, either traditional (i.e. citation count) or alternative (i.e. altmetrics). On the contrary, the experiment based on machine learning assumes metrics as independent features. In fact, it investigates which metrics are effective for evaluating scholars, that is to predict those who are eligible for the scientific qualification by using bibliometric indicators. We combined 12 different configurations of metrics used as classification features. The rationale associated with the configurations is the following:

\begin{figure}[!ht]
\centering
\subfigure[Full Professor.]{\includegraphics[width=0.8\textwidth]{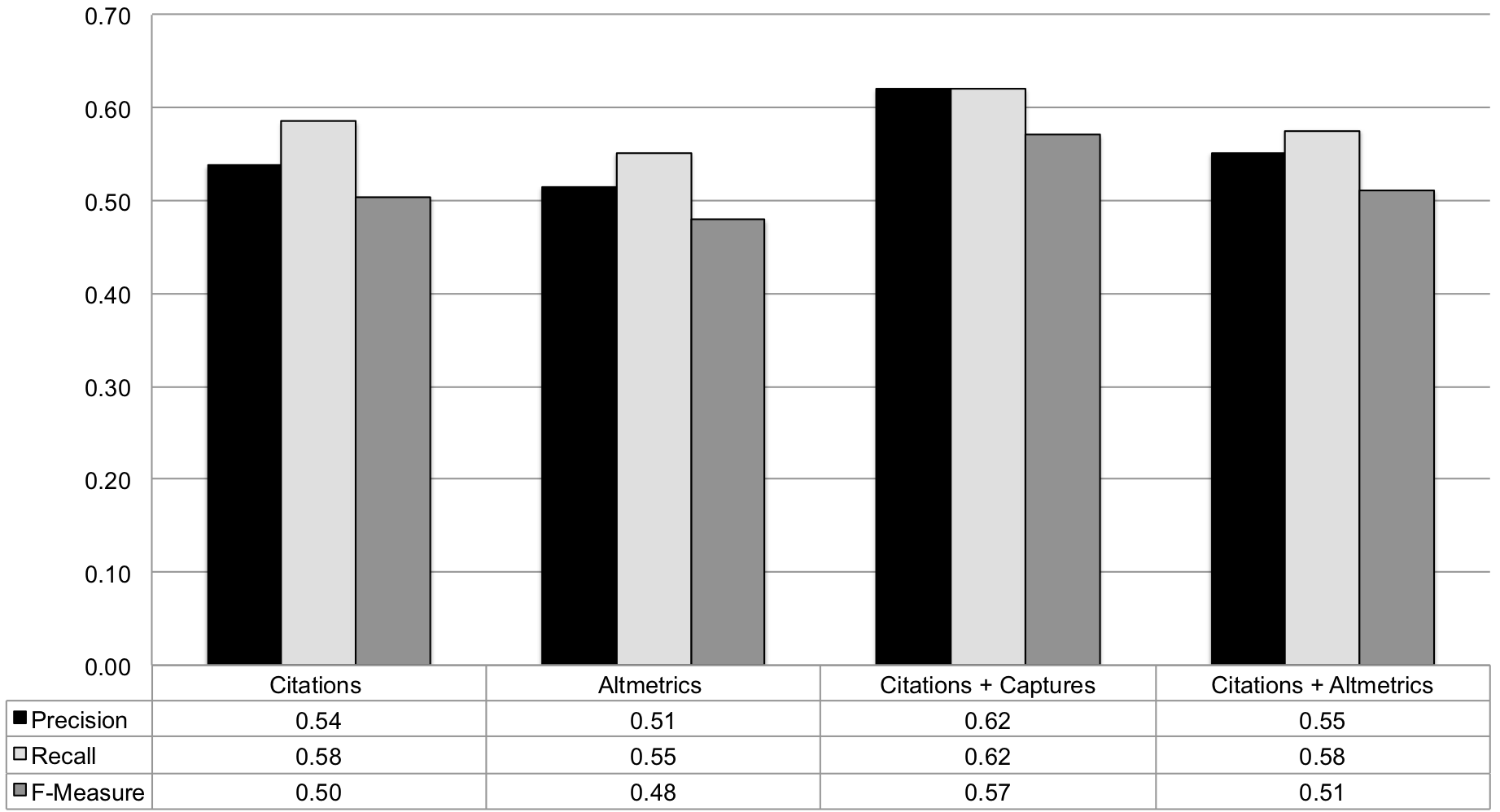}}\label{fig:classification-count-level-1}
\subfigure[Associate Professor.] {\includegraphics[width=0.8\textwidth]{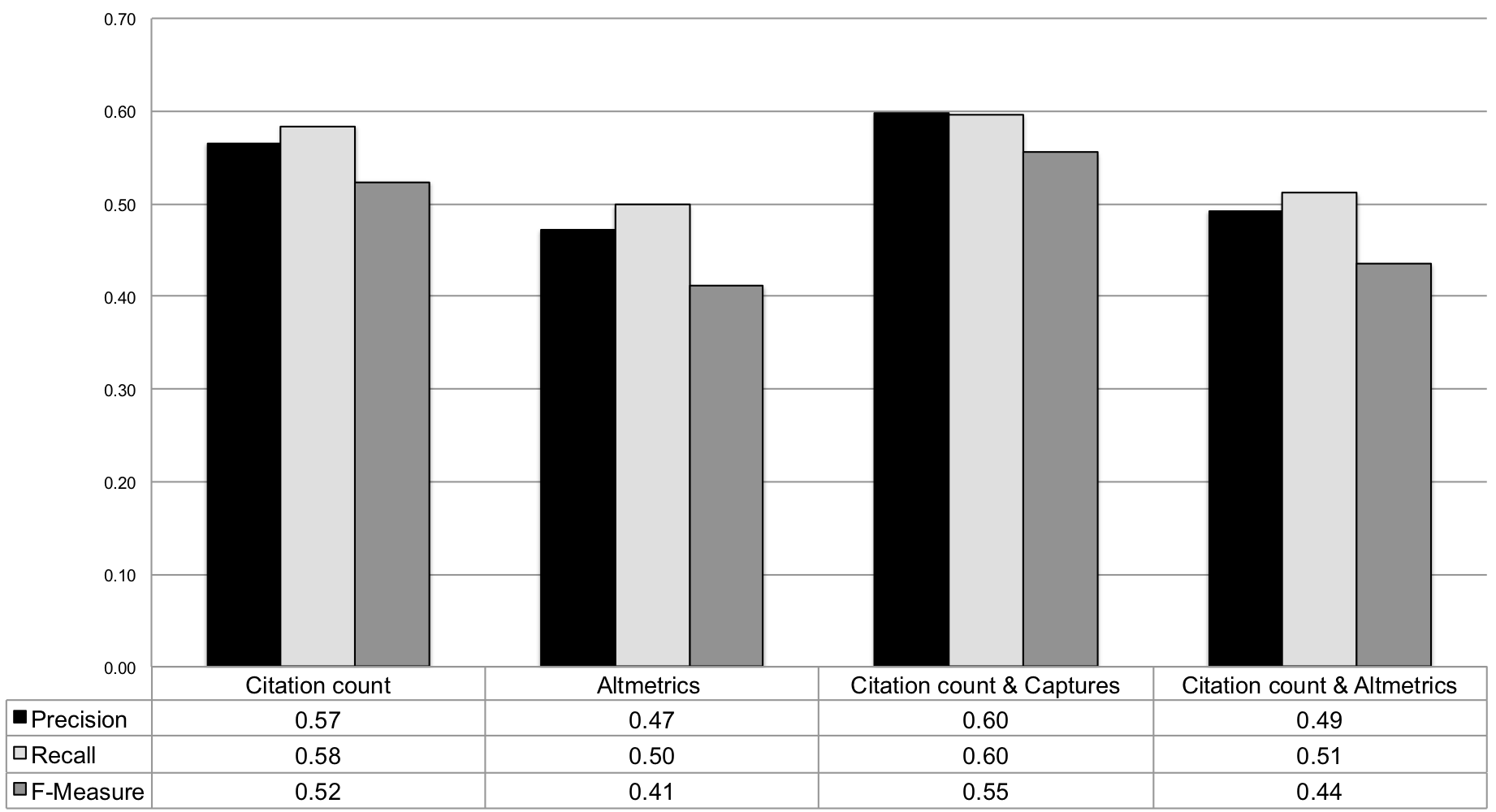}}\label{fig:classification-count-level-2}
\caption{Precision, recall, and accuracy computed by using article level indicators as features.} \label{fig:classification-count-levels}
\end{figure}

\begin{itemize}
\item we identified 3 sets of indicators: (i) citation counts only, (ii) captures only, and (iii) the whole set of altmetrics categories. We selected the citation counts because it currently identifies the most adopted indicator used for measuring the impact of research. Similarly, we selected the captures as, according to the analysis of the correlation coefficients (cf. Section~\ref{sec:results-corr}), it closely mirrors citation count. Finally we selected the whole set of the categories of altmetrics because they are the principal object of our investigation.
\item we generated different combination of the 3 sets of metrics. However, we never combine captures and the whole set of altmetrics as the latter already includes the former. This produced 4 different configurations of features to use with the Na\"{i}ve Bayes classifier. Namely, the configurations are the following: (i) citation counts only, (ii) altmetrics only, (iii) citation counts and captures, and (iv) citation counts and altmetrics;
\item we had other 4 configurations resulting from the $h$-index indicators, i.e. (i) $h$-index based on citation counts only, (ii) $h$-index based on altmetrics only, (iii) $h$-index based on citation counts and $h$-index based on captures, and (iv) $h$-index based on citation counts and $h$-index based on altmetrics;
\item we had other 4 configuration resulting by combining article level indicators and author level indicators, i.e. (i) citation counts and $h$-index based on citation, (ii) altmetrics and $h$-index based on altmetrics, (iii) citation count, $h$-index based on citation count, captures, and  $h$-index based on captures, and (iv) citation count, $h$-index based on citation count, altmetrics, and $h$-index based on altmetrics.
\end{itemize}

\begin{figure}[!ht]
\centering
\subfigure[Full Professor.]{\includegraphics[width=0.8\textwidth]{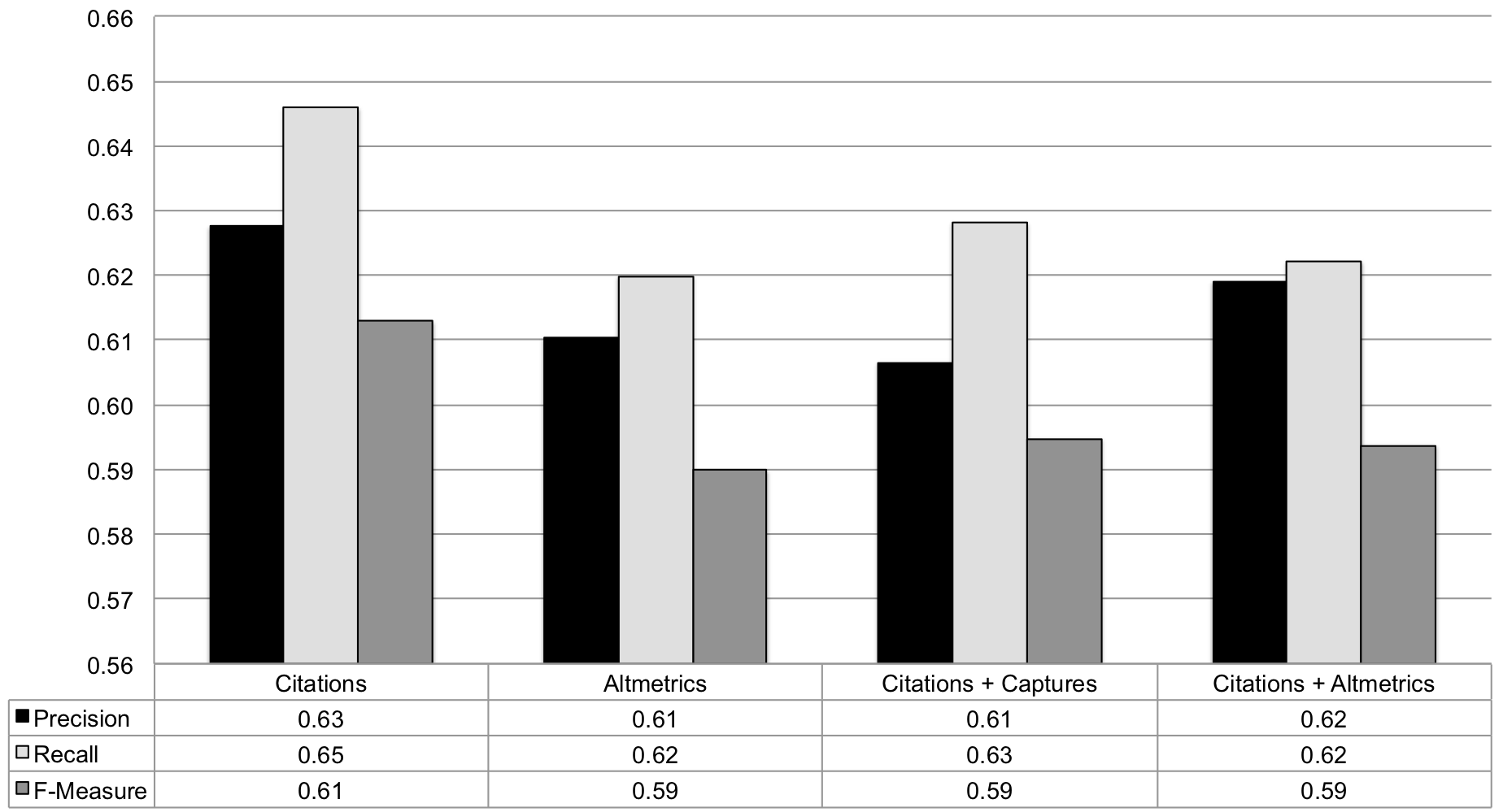}}\label{fig:classification-h-index-level-1}
\subfigure[Associate Professor.] {\includegraphics[width=0.8\textwidth]{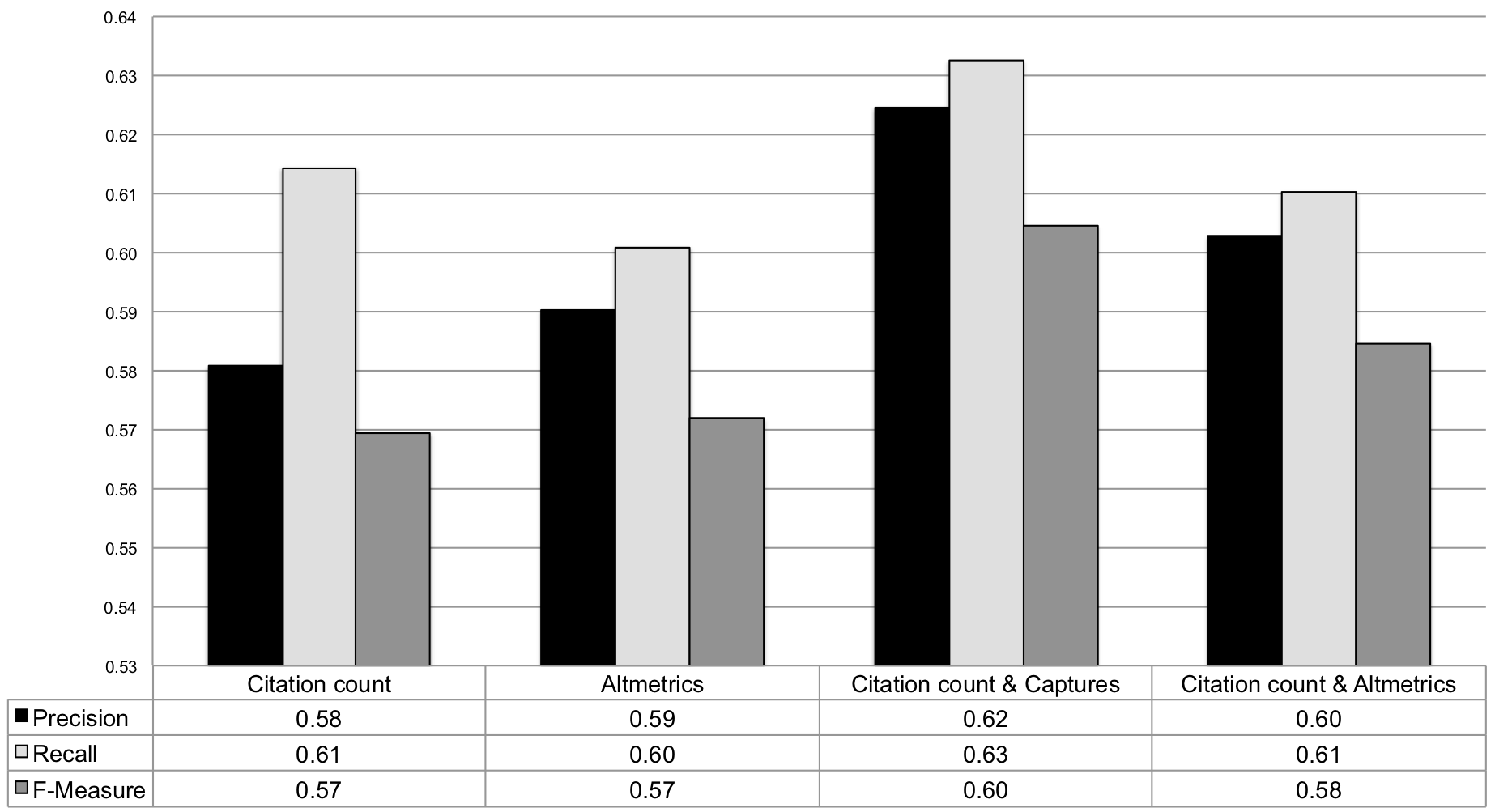}}\label{fig:classification-h-index-level-2}
\caption{Precision, recall, and accuracy computed by using h-indexes based on citation counts and altmetrics as features.} \label{fig:classification-h-index-levels}
\end{figure}

Figure~\ref{fig:classification-count-levels} shows the results in terms of precision, recall, and F-Measure when the article level indicators, consisting of the citation counts and almetrics, are used as features. The results are reported both for the full professor level (cf. Figure~\ref{fig:classification-count-level-1}) and the associate professor level (cf. Figure~\ref{fig:classification-count-level-2}). With this set of features we observe that the N\"{a}ive Bayes algorithm performs better when the citation counts are used together with the captures. In fact, with this configuration we recorded $precision=0.62$, $recall=0.62$, and $F$-Measure=0.57 for the full professor level and $precision=0.6$, $recall=0.6$, and $F$-Measure=0.55 for the associate professor level.

\begin{figure}[!ht]
\centering
\subfigure[Full Professor.]{\includegraphics[width=0.8\textwidth]{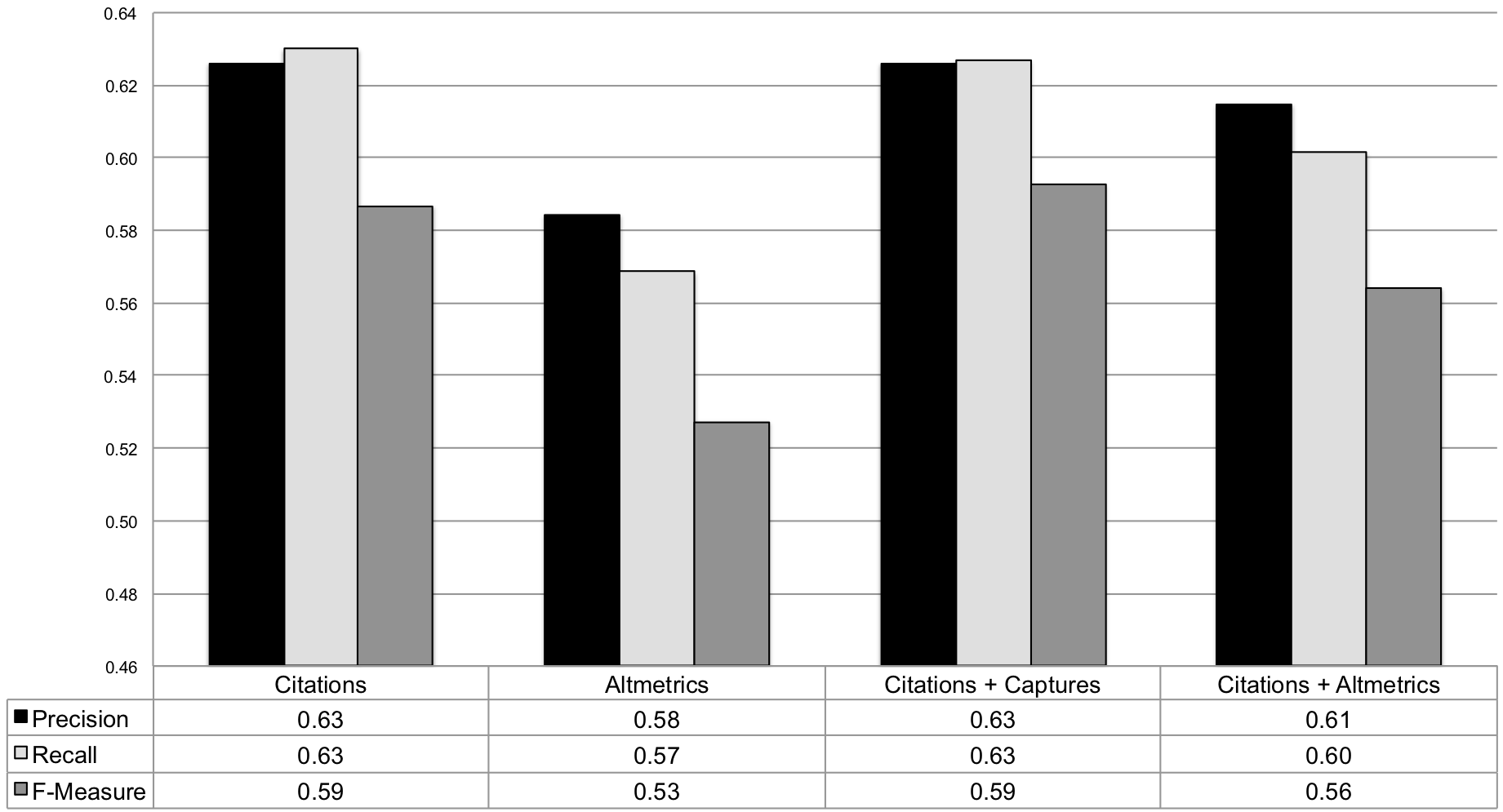}}\label{fig:classification-count_h-index-level-1}
\subfigure[Associate Professor.] {\includegraphics[width=0.8\textwidth]{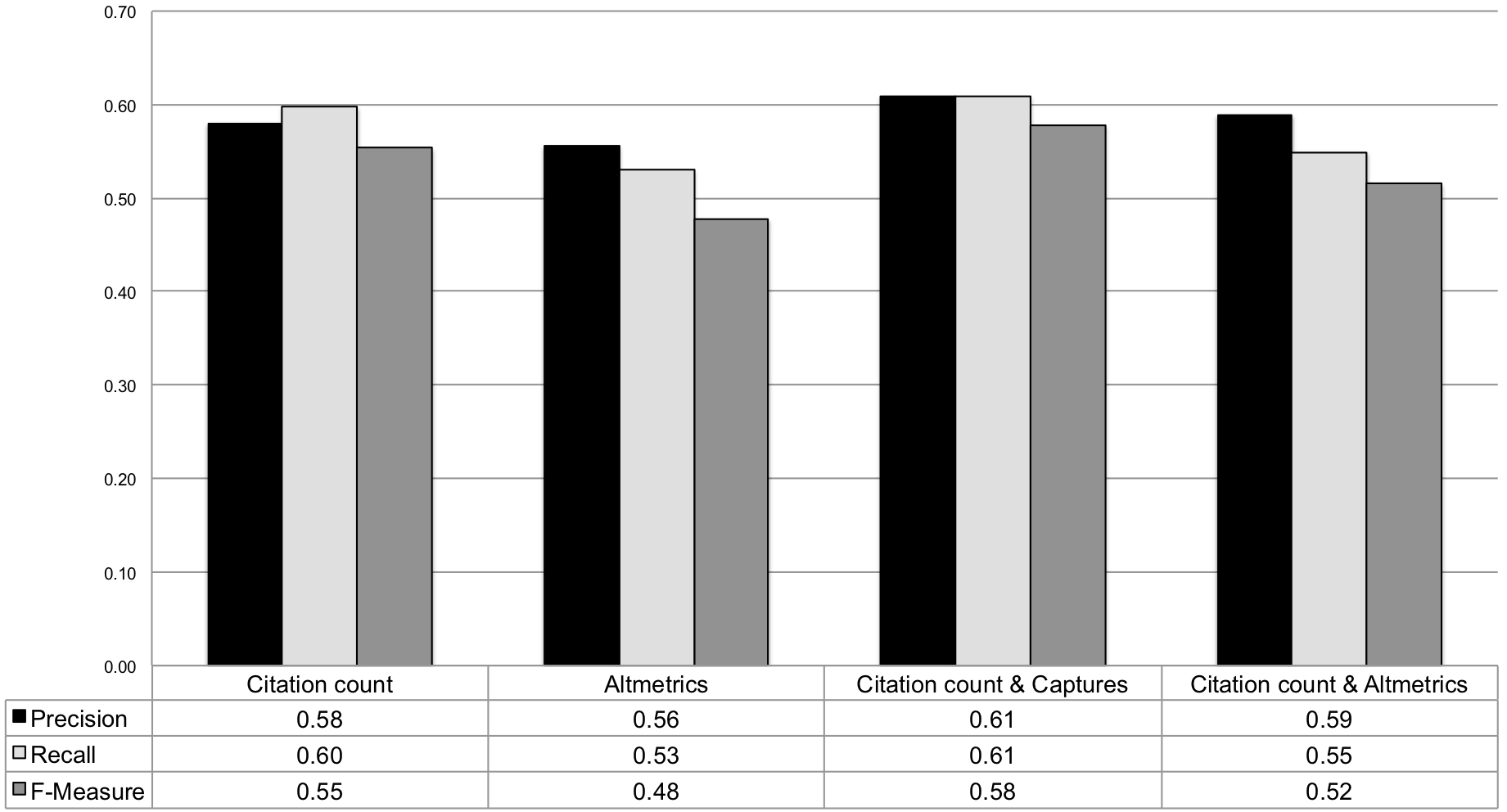}}\label{fig:classification-count_h-index-level-2}
\caption{Precision, recall, and accuracy computed by using both article level indicators and author level indicators as features.} \label{fig:classification-count_h-index-levels}
\end{figure}

Figure~\ref{fig:classification-h-index-levels} shows the results in terms of precision, recall, and F-Measure when the $h$-indexes obtained from the citation counts and almetrics are used as features. The results are reported both for the full professor level (cf. Figure~\ref{fig:classification-h-index-level-1}) and the associate professor level (cf. Figure~\ref{fig:classification-h-index-level-2}). In the case of the full professor level we observe that the N\"{a}ive Bayes algorithm performs better when it is trained by using the $h$-indexes computed on citation counts as features, i.e. $precision=0.63$, $recall=0.65$, and $F$-Measure=0.61. Instead, in the case of the associate professor level we observe better performance when the algorithm is trained by using the $h$-indexes based both on the citation counts and the captures as feature, i.e. $precision=0.62$, $recall=0.63$, and $F$-Measure=0.60.

Finally, Figure~\ref{fig:classification-count_h-index-levels} shows the results in terms of precision, recall, and F-Measure when both article level indicators and author level indicators are used as features. Again, the figure reports the results both for the full professor level (cf. Figure~\ref{fig:classification-count_h-index-level-1}) and the associate professor level (cf. Figure~\ref{fig:classification-count_h-index-level-2}). In this latter configuration we observe, for the full professor level, that the algorithm performs better in either cases it is trained by using as features the citation counts only or the citation counts along with the captures, i.e. $precision=0.63$, $recall=0.63$, and $F$-Measure=0.59. On the contrary, for the associate professor level, the algorithm performs better when the algorithm is trained by using as feature both the citation counts and the captures, i.e. $precision=0.61$, $recall=0.61$, and $F$-Measure=0.58.


\subsection{Discussion}
The correlation analysis shows that the citation counts and the captures category of altmetrics as well as the social media and the mentions categories correlate better than other pairs of indicators.
Although the correlation between these pairs is only moderate (i.e. $r$=$\sim$0.55 and $r$=$\sim$0.5 for ``citation count-captures'' and ``social media-mentions", respectively), its statistical significance is high according to $p$-values computed on the Student's $t$-distribution (i.e. $p<0.01$). 
If we move the analysis forward to the level of the metrics we observe that the moderate correlation between citation counts and captures is fed by the number of readers, which is one of the metric part of the captures category. In fact, we recorded a moderately high and statistically significant ($r$=$\sim$0.6 and $p<0.01$) correlation when the number of readers is compared to the citation counts. According to our findings, the latter moderately high correlation is recorded when Mendeley is used as source for the number of readers (i.e. $r$=$\sim$0.6 and $p<0.01$). This findings is not surprising as, according to different studies in literature~\cite{Costas2015,Ortega2018}, Mendeley is the primary source for altmetrics in PlumX.
Instead, if we investigate on the correlation between social media and mentions we observe that tweets on Twitter and posts on Facebook for social media, and blogs for mentions play a relevant role in generating the moderate correlation recorded (i.e. $r$=$\sim$0.45). This moderate correlation, which is statistical significant (i.e. $p<0.01$), seems to confirm a shared intended use of social media and blogs based on microblogging for enabling scholarly communication. As a matter of fact, (i) Twitter is the most widespread microblogging service; (ii) Facebook provides microblogging capabilities by means of status updates; and (iii) many researchers use microblogging (i.e. frequent and short posts) on their own personal blogs as well as on community based ones either to boost their scientific impact or, more in general, for scholarly communication. The author perspective shows a slightly different scenario. In fact, (i) the correlation coefficients are more homogeneously spread and (ii) we observe good correlations in a few cases, which are significantly higher than those recorded for article level indicators. Namely, those cases are the ``citation count-captures'' ($r$=$\sim$0.78) and the ``citation count-usage'' ($r$=$\sim$0.65) pairs. In both cases we recorded the correlation as statistically significant (i.e. $p<0.01$). However, while a moderate correlation between the citation counts and the captures is observed at article level, the same does not hold for the correlation between the citation counts and the usage. In fact, we do not observe any correlation between the two indicators at the article level (i.e. $r$=$\sim$0.1). This suggests three alternative speculations, i.e. (i) the comparison of indicators under the author assumption is more reliable than that performed under the article assumption; (ii) the author assumption (more specifically, that based on the $h$-index) introduces a bias that affects the comparison negatively or, viceversa, the article assumption does the same; and (iii) the correlation analysis is reliable in both cases as the author assumption and the article one provide different insights on the indicators. Our intuition is that they are both reliable.
This intuition is confirmed by the second experiment based on the automatic classification of scholars in terms of qualified or not qualified to one of the two professional levels of the Italian NSQ. In this experiment we defined three different conditions based on three different configuration of the set of features, i.e. (i) we used only article level indicators; (ii) we used only author level indicators; and (iii) we used both levels. According to our results, none of the configurations outperforms the others significantly. In fact, they all provide comparable results in terms of precision, recall, and accuracy. We only observe that the classification algorithm performs slightly better when the $h$-indexes are used as features either alone or combined with article level indicators. However, this outcome is feasible as the goal of the qualification procedure is to evaluate scholars. Hence, the author level indicators are in general more appropriate for this kind of assessment. We remark that the performance of the classification (i.e. $\sim$0.55 as $F$-Measure) should not misinterpreted. In fact, the committees of peers use a plethora of heterogeneous criteria that can be gathered from the curricula besides those based on traditional indicators like the citation counts and the $h$-index. On the contrary, the $F$-Measure recorded is surprisingly high if take into consideration that the automatic classification algorithm emulates humans by using traditional indicators and altmetrics only. Our experiment is not intended at demonstrating that automatic systems can substitute committees of peers in academic evaluation procedures. Nevertheless, we believe that automatic systems can provide effective support during evaluation procedures, like the Italian NSQ, to both the committees of peers and the candidates. For example, a candidate might ask an automatic system to assess if she is in a good position for the qualification. Similarly, a committee of peer might use an automatic system to filter out clear true negatives from the bunch of applications and shorten the response time frame. Finally, we did not take into account self-citations and gaming that, in the case of altmetrics, can be even more frequent and unfavorable. In fact, the time horizon of altmetrics is shorter and, in most of the cases, their trust is limited due to their decentralisation with respect to the publishing authority. It is obvious that gaming can affect the results of the automatic prediction. In general this is valid also for peer-based assessment procedures, though peers try to filter out self-citations and gaming by human judgment. However, on one hand, the prediction algorithm is trained on the results of the evaluation produced by human peers, then it is supposed to learn from their examples. On the other hand, we believe that the introduction of automatic methods for detecting and filtering self-citations and gaming can produce better predictions.

\section{Conclusions and future work}
\label{sec:conclusion}
In this work we investigate the relevance and the effectiveness of alternative indicators when used for assessing research quality. First, we provide a background about altmetrics with their definition (cf. Section~\ref{sec:background}). Then, we describe the material used for experimenting with altmetrics in order to analyse their effectiveness. Namely, we use the curricula available from the first National Scientific Qualification (NSQ) held in Italy in 2016 (cf Section~\ref{sec:data}). The data from those curricula are used for populating an RDF knowledge graph, named MIRA-KG, which counts of $\sim$21M triples (cf. Section~\ref{sec:experiment-datasample}). Such a knowledge graph, modelled in compliance with the Indicators Ontology (i.e. I-Ont), serves as the playground for our experiments (cf. Section~\ref{sec:experiment-datasample}), which are, namely:
\begin{itemize}
\item the analysis of the correlations between indicators, either traditional (i.e. citation count and $h$-index) or alternative (i.e. altmetrics);
\item the automatic prediction of the candidates' qualification at the NSQ by using independent settings of indicators as features for training a N\"{a}ive Bayes algorithm.
\end{itemize}
The first experiment investigates what are the group of indicators that share a common intended meaning. The second one investigates which indicators more effectively mirror humans in the task of evaluating scholars. For the latter experiment we use the results produced by the committees of peers of the NSQ as groundtruth. According to the results the number of readers on Mendeley is recorded as the source of alternative metrics that best correlate with traditional metrics both at the article level (i.e. citation count) and author level (i.e. $h$-index). This is compliant with literature~\cite{Bornmann2018altmetrics,Thelwall2018}. Furthermore, Mendeley reads are also recorded as the metric that provides more accurate (in terms of $F$-measure) results for prediction of candidates' qualification to the NSQ.
Future work includes the extension of the experiments to the whole dataset of the NSQ by possibly including data from sessions following the first of 2016. Other possible future work is the adoption of more comprehensive sets of bibliometric and non-bibliometric indicators for experimenting with our methodology.

\section*{Acknowledgements}
This research has been supported by the Italian National Agency for the Evaluation of the University and Research Systems (ANVUR) within the Measuring the Impact of Research - Alternative indicators (MIRA) project.

\bibliographystyle{aps-nameyear}      
\bibliography{references}                 
\nocite{*}

\end{document}